\documentclass [twocolumn,aps,prb,showpacs]{revtex4}
\usepackage{graphicx,amssymb,color}
\pdfoutput=1

\begin{document}
\title{An exciton scattering model for carrier multiplication in semiconductor nanocrystals: Theory}
\author{Andrei Piryatinski}\email{apiryat@lanl.gov}
\author{Kirill A. Velizhanin}
\affiliation{Center for Nonlinear Studies (CNLS), Theoretical Division,
Los Alamos National Laboratory, Los Alamos, NM 87545 }
\date{\today}

\begin{abstract}
 The effect of carrier multiplication (CM) in semiconductor nanocrystals is systematically treated by 
 employing an exciton scattering approach. Using projection operators, we reduce the Coulomb 
 coupled multi-exciton dynamics to scattering dynamics in the space spanning both single- and 
 bi-exciton states. We derive a closed set of equations determining the scattering matrix elements.
 This allows us to interpret CM dynamics as a series of odd-order interband scattering events. Using the 
 time-dependent density matrix formalism, we provide a rigorous description of the CM dynamics induced by a finite-time 
 pump pulse. Within this approach, both processes of single- and bi-exciton photogeneration and the consequent 
 population relaxation are treated on the same footing. This approach provides a framework for numerical calculations
 and for comparisons of the quantum efficiencies associated with each process. For applications,
 the limit of weak interband Coulomb coupling is considered. Finally, we demonstrate that three previously used
 theoretical models can be recovered as limiting cases of our exciton scattering model. 
\end{abstract}

\maketitle


\section{Introduction}
\label{intro}

Carrier multiplication (CM) in semiconductor materials is the process of more than one electron-hole pair generation per single absorbed photon. Here, we consider the general case in which high energy electron-hole pairs consist of free carriers (as typically occurs in bulk semiconductors). We also consider the case in which the carriers are confined exciton states as occurs in semiconductor nanocrystals (NCs). CM is naturally characterized by the related Quantum Efficiency (QE) which is the number of electron-hole pairs generated per absorbed photon. CM is also characterized by the activation energy threshold (AET) below which CM becomes negligible. Extensive studies of CM are motivated by potential applications in photovoltaic, photoelectrochemical, and energy storage devices.\cite{kolodinski93,nozik02,hanna06,klimov06,beard08a,nozik08,luther08} 

CM was first investigated in bulk materials using photocurrent measurements,\cite{koc57,smith58,vavilov59,tauc59,cristensen76} and was recently revisited using terahertz time-domaing spectroscopy.\cite{pijpers09} The theory of CM in bulk treats CM as a sequence of the primary photoexcitation event in which a {\em single} electron-hole pair is created by a photon, and the secondary process of the electron and hole population relaxation during which CM occurs.\cite{shockley61a,kane67,antoncik78,wolf98} The population relaxation dynamics is a competition between the impact ionization process in which the excess kinetic energy of the hot electron or hole is transferred to create another electron-hole pair\cite{kane67,landsberg91,harrison99} and the process of phonon-assisted cooling.\cite{antoncik78,wolf98}    
   
In bulk, strict energy {\it and} quasi-momentum conservation constraints determine the values of AET\cite{klein68,alig75} and QE as a function of the absorbed photon energy\cite{antoncik78}. For a variety of semiconductor materials, the lower boundary of AET is found to be about $3E_g$ where $E_g$ is the bulk band gap energy.\cite{klein68,alig75} However, photocurrent \cite{koc57,smith58,vavilov59,tauc59,cristensen76} and optical \cite{pijpers09} measurements have demonstrated that the AET for most materials is $\gtrsim 4E_g$.\cite{koc57,smith58,vavilov59,tauc59,cristensen76} 

In semiconductor NCs, it is expected that the following three processes should lead to an increase in QE and a decrease in AET:  relaxation of the quasi-momentum conservation constraint\cite{chepic90}, a decrease in the phonon-assisted relaxation rate\cite{nozik01}, and an enhancement of Coulomb interaction between the carriers\cite{chepic90,klimov00}. Efficient CM has been reported in colloidal NCs using time-resolved transient absorption (pump-probe) and time-resolved photoluminescence techniques.\cite{schaller04,ellingson05,schaller05,schaller05a,schaller06,schaller06a,schaller06b,murphy06,schaller07,beard07,pijpers07} Reported values of AET vary in the range of $2-3E_g$ with $E_g$ being the NC's band gap energy. By using a bulk-type model with relaxed quasi-momentum conservation rule, it has been further speculated that, depending on the ratio of electron and hole effective masses, the AET can reach a minimum value of $2E_g$ satisfying the energy conservation constraint.\cite{schaller05a}   

These experimental results have been challenged by a number of reports claiming significantly lower QE and even the absence of the CM effect.\cite{nair07,benlulu08,pijpers08} CM has been further reconfirmed, however, with observed values of QE varying in a broad range starting below the QE in bulk materials.\cite{mcguire08,trinh08,nair08,ji09} The variation of QE could possibly arise from experimental inaccuracies,\cite{benlulu08,franceschetti08} sample-to-sample variation in surface preparation,\cite{beard09,kilina09a} and extraneous effects such as photocharging\cite{mcguire08}. These issues raise an important question: What are the specific quantum-confinement-induced features that distinguish CM in NCs from CM in bulk semiconductors?\cite{nair08} Addressing this question requires theoretical insight. Currently, there are three separate models outlined below proposing different mechanisms for CM in NC: the Coherent Superposition Model, the Direct Photogeneration Model, and the Impact Ionization Model. We describe each of these in the following few paragraphs.   

{\it The Coherent Superposition Model} of resonant (almost degenerate) single- and bi-exciton states is based on the density matrix formalism. It was proposed by Shabaev, Efros, and Nozik.\cite{shabaev06} This model states that, in contrast to bulk materials, the primary event of single photon absorption in NCs leads to the preparation of coherent superpositions (oscillations) between the single- and bi-exciton states that are almost degenerate. The secondary process of phonon-induced intraband relaxation merely stabilizes the populations leading to efficient bi-exciton production due to the fast bi-exciton intraband relaxation rate. No experimental observations of these oscillations have been reported yet. The enhancement of QE according to this model requires a strong Coulomb coupling between single- and bi-exciton states. This enhancement has not been confirmed experimentally. This model ignores the effects of the single-/bi-exciton density of states (DOS) by considering only one single-exciton and one bi-exciton states coupled through Coulomb interactions. As we demonstrate in this paper, this model also misses the CM pathway that involves the phonon-assisted relaxation channel between single- and bi-exciton states. 

Assuming weak Coulomb coupling between single- and bi-exciton states and assuming optical pulse duration larger than the dephasing time, the QE can be evaluated using Fermi's Golden Rule.\cite{schaller05,rupasov07} This approach, referred as the {\it Direct Photogeneration Model}, predicts two pathways for direct bi-exciton production during the primary photon absorption event. The first pathway, introduced by Schaller, Agranovich and Klimov, describes resonant bi-exciton generation via virtual single-exciton states.\cite{schaller05} The second pathway, considered by Rupasov and Klimov, accounts for the non-vanishing Coulomb matrix elements between the exciton vacuum (filled valence band) and bi-exciton states. This coupling leads to the stabilization of bi-exciton populations through resonant intraband optical transitions.\cite{rupasov07} 

These authors estimate the contributions of their respective pathways and claim that their pathways become efficient in NCs because the quasi-momentum conservation constraint is relaxed. The actual enhancement of QE comes from the increased bi-exciton DOS compared with the single-exciton DOS. Independent quantum chemistry calculations confirm the possibility of direct carrier photogeneration in semiconductor clusters.\cite{isborn08} The drawback of the Direct Photogeneration Model is that no secondary events of population relaxation on QE are considered. In our paper, we will also demonstrate that the additional channel associated with the direct excitation of single-exciton states and their further scattering to the bi-exciton manifold during the interaction with the optical pulse as well as the interference of all the pathways must be included in the weak Coulomb limit.

A number of reported calculations suggest that, in contrast to the mechanisms outlined above and similar to the bulk materials, CM in NCs occurs solely due to the competing phonon-assisted relaxation and impact ionization processes that follow the primary single-exciton photoexcitation event. We will refer to this approach as the {\it Impact Ionization Model} model throughout this paper. Specifically, Franceschetti, An, and Zunger have considered the spectral dependence of the impact ionization rate and Auger recombination (the inverse process) rate using atomistic pseudopotential calculations.\cite{franceschetti06} Allan and Delerue used a tight-binding model to simulate the competing processes of impact ionization and phonon-assisted relaxation.\cite{allan06} Their analysis based on their models emphasizes the importance of the high ratio of bi- to single-exciton DOS for efficient CM. 

Further development of this approach led to a DOS-based comparison of QE due to impact ionization and direct photogeneration,\cite{allan08} to evaluation of the band-structure effects on QE in a variety of NC materials,\cite{luo08} and to modeling the influence of surface defects on QE.\cite{allan09} Interestingly, Rabani and Baer emphasized the importance of the trion DOS (in contrast to the bi-exciton DOS) directly entering the impact ionization and Auger recombination rates, where strict selection rules enter through the Coulomb matrix elements.\cite{rabani08}    

Currently, the Coherent Superposition Model, the Direct Photogeneration Model, and the Impact Ionization Model are considered as alternative approaches whose applicability is still being debated. We propose a more general approach capable of treating the CM dynamics in both NCs and bulk materials by accounting for both the photogeneration event induced by a finite-time optical pulse and the population relaxation dynamics, on the same footing. This approach can be used to interpolate between strong and weak Coulomb coupling regimes. This interpolation can be achieved by treating Coulomb interactions between the carriers as multiple-scattering events. We achieve this more general approach by integrating the scattering theory with the density matrix formalism, and we call this approach the {\it Exciton Scattering Model}. 

As a validation of our Exciton Scattering Model, we demonstrate that the previously proposed models can be recovered as limiting cases, and they are fundamentally related to each other. We also demonstrate that the proposed model predicts additional contributions to the Coherent Superposition and the Direct Photogeneration models which have not been considered before. We use our Exciton Scattering Model to formulate a closed computational scheme for determining QE and ATE. The results of the numerical simulations using this approach applied to specific semiconductor materials will be reported in a separate paper.

This paper is organized as follows: In Sec.~\ref{xdhs}, the projection operator technique is employed to reduce the coupled multi-exciton dynamics to single- and bi-exciton scattering dynamics in Hilbert space. We use this technique to derive a closed set of equations for the interband scattering matrix. In Sec.~\ref{prdl}, we use the density matrix formalism combined with a modified exciton scattering approach to obtain general expressions for the QE that naturally describe the primary event of single- and bi-exciton photogeneration due to both a {\em finite}-time pump pulse and population relaxation dynamics. For numerical calculations, the limiting case of weak Coulomb coupling is introduced in Sec.~\ref{wclm} in which a closed set of equations for the limiting QE is presented. In Sec.~\ref{disc}, we discuss connections of the Exciton Scattering Model with previously proposed models. Finally, we present our conclusions in Sec.~\ref{conc}.   

\section{Milti-exciton dynamics in Hilbert space}
\label{xdhs}

In this section, we begin our analysis by introducing the many-body electronic Hamiltonian in the multi-exciton representation accounting for the contributions of all of the Coulomb terms. These terms can be partitioned into those terms that conserve the total number of excitons (and determine their binding energies), and those terms that do not conserve the number of excitons, giving rise to the CM dynamics. An exact treatment of the dynamics of total multi-exciton space is not feasible. Therefore, we restrict our dynamics to the reduced space spanning  single- and bi-exciton states by using the projection operator technique. This approach allows us to include some of the effects of higher-multiplicity (tri-, four-, {\it etc.}) exciton states in the dynamics in the reduced space. Since the projected dynamics is restricted to coupled single- and bi-exciton manifolds only, it can be treated by performing an exact summation of the perturbation series, in which the {\it odd-order} interband scattering events describe CM dynamics.

\begin{figure}[t]
\centerline{\includegraphics[width=3.5in]{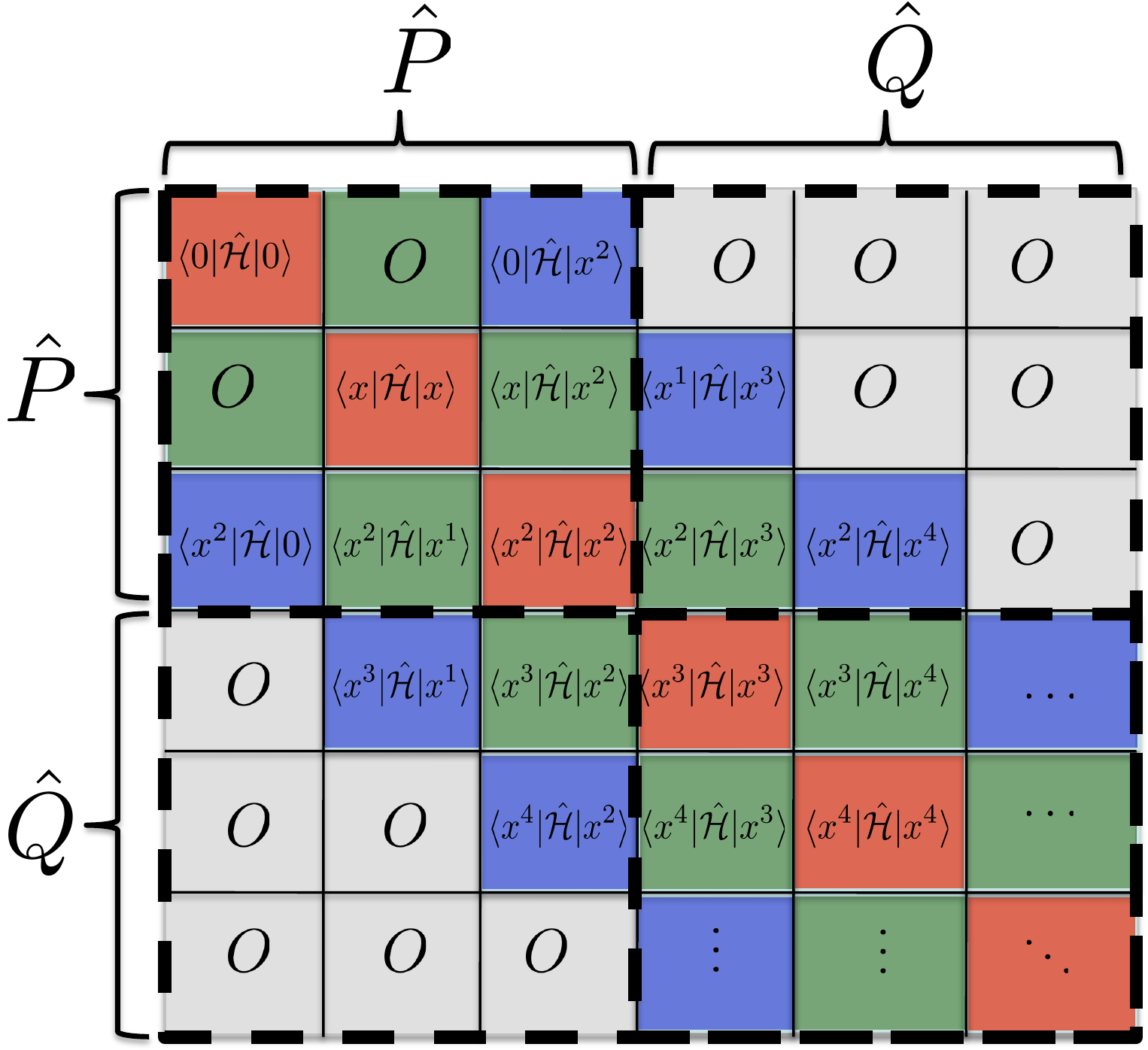}}
    \caption{The multi-exciton Hamiltonian in the block-matrix representation. The $\langle 0|\hat{\cal H}|0\rangle$-block is the exciton vacuum, and the rest of the diagonal (red) blocks are the Hamiltonian components in the single-exciton, bi-exciton, {\it etc.} subspaces. The green off-diagonal blocks describe the Coulomb interactions between the latter components changing exciton multiplicity by one, and the blue blocks by two. $0$-blocks indicate null-matrices. Four quadrants separated by the black dashs describe the partitioning of the multi-exciton Hamiltonian by the projection operators $\hat P$ and $\hat Q$. The upper left quadrant is the projected Hamiltonian (Eqs.~(\ref{Hs}) -- (\ref{HW})) acting in the space spanning the single- and bi-exciton states.}
    \label{Fig-Hmlt}
\end{figure}

\subsection{The multi-exciton Hamiltonian}
\label{mexh}

Let us consider the valence and conduction bands of a semiconductor NC in which single electron and hole states are known within the Hartree-Fock approximation (or equivalently within the effective mass envelope function formalism).  Our many-body electronic Hamiltonian, $\hat{\cal H}$, accounts for these non-interacting single particle states, and all possible Coulomb interactions among them.\cite{axt98} An explicit form of this Hamiltonian is given in Appendix~\ref{EffH}. Not all the Coulomb interaction terms in $\hat{\cal H}$ conserve the total number of electrons and holes. However, this Hamiltonian does conserve total charge. Consequently, the dynamics of electrically neutral electron-hole pairs (excitons) is uncoupled from the dynamics of the charged states.  This allows us to focus on the dynamics determined by the multi-exciton Hamiltonian,
\begin{eqnarray}\label{HX}
    \hat H_{MX} = \hat H_{MX}^{(0)} + \hat H_{MX}^{(1)} + \hat H_{MX}^{(2)},
\end{eqnarray}
whose derivation is provided in Appendix~\ref{EffH}.

The block-matrix representation of this multi-exciton Hamiltonian is shown in Fig~\ref{Fig-Hmlt}, in which the $\langle 0|\hat{\cal H}|0\rangle$-block denotes the exciton vacuum, i.e. the filled valence band, with its energy set to zero. The remaining diagonal (red) blocks describe the single-exciton, bi-exciton, {\it etc.}, sub-spaces. These terms are:
\begin{eqnarray}\label{HX0}
    \hat H_{MX}^{(0)} &=& \sum_{{\bar n\geq1}}\sum_{ p\geq1} | x_{p}^{{\bar n}}\rangle \hbar\omega^{\bar n}_p
    \langle x_{p}^{\bar n}|,
\end{eqnarray}
in which $| x_{p}^{\bar n}\rangle$ denotes the $p$-th exciton state with multiplicity, ${\bar n}$, and energy, $\hbar\omega^{\bar n}_p$. This energy already includes the ${\bar n}$-particle binding interactions which can be calculated, e.g. by block-wise matrix diagonalization.

The off-diagonal Coulomb interaction (green and blue) blocks do not conserve the total number of electrons and holes. They describe the interactions between exciton states with different multiplicity. Specifically, the green off-diagonal blocks describe the processes changing multiplicity by one. The blue off-diagonal blocks change multiplicity by two. Note that the bi-exciton states are coupled to the vacuum, whereas the single-exciton states are uncoupled from the vacuum. This is a result of the Hartree-Fock representation eliminating the latter interactions. The general expression for these off-diagonal terms in the multi-exciton Hamiltonian is
\begin{eqnarray}\label{HX12}
    \hat H_{MX}^{(\bar i)} &=& \sum_{\bar n} \sum_{pq}
     |x_{p}^{\bar n}\rangle V^{{\bar n},{\bar n}+\bar i}_{p,q} \langle x_{q}^{{\bar n}+\bar i}|
    + h.c.,
\end{eqnarray}
in which $V^{{\bar n},{\bar n}+\bar i}_{p,q}=\langle x_{p}^{\bar n}|\hat{\cal H}|x_{q}^{\bar{n}+\bar i}\rangle$ is the interband multi-exciton interaction matrix element with $\bar i=1,2$ describing the multiplicity variation.

\subsection{Projected dynamics in single- and bi-exciton space}
\label{rdyn}

The dynamics in the total multi-exciton Hilbert space is fully defined by the propagator:
\begin{eqnarray}\label{U-MX}
    \hat U (t) = \Theta(t)\exp\left[-i\hbar^{-1}\hat H_{MX}t\right],
\end{eqnarray}
whose calculation and general analysis is not feasible due to the rapidly growing number of multi-exciton states. Thus, we introduce a convenient representation allowing us to approximate calculations of this propagator.

The CM processes excited near the AET should primarily result in photogeneration of single- and bi-exciton states. On the other hand, the higher-multiplicity exciton states can still affect their dynamics. Therefore, we consider a dynamics projected onto the space spanned by the single- and bi-exciton states, and seek the conditions allowing us to neglect the effects of the higher-multiplicity states. This can be naturally done by introducing the following projection operator onto the space of single- and bi-exciton states,
\begin{eqnarray}\label{PQ-oper}
    \hat P = \sum_{a\geq0}|x_a\rangle\langle x_a|
    			+ \sum_{k\geq1}|xx_k\rangle\langle xx_k|,
\end{eqnarray}
and the complementary projection operator,  $\hat Q = \hat I -\hat P$ onto the rest of the multi-exciton space. $\hat I$ denotes the identity operator  in total multi-exciton space. To distinguish the single- and bi-exciton states from the remaining higher-multiplicity states, $|x_{p}^{\bar n}\rangle$ where $n \geq 3$, we have introduced their new notations $|x_a\rangle$ and $|xx_k\rangle$, respectively, and use $|x_0\rangle$ to denote the exciton vacuum. This notation will be  used through out this paper.

The partitioning of the multi-exciton Hamiltonian by the projection operators, $\hat P$ and $\hat Q$, is illustrated in Fig.~\ref{Fig-Hmlt}, where the {\em projected} Hamiltonian $\hat H =\hat P\hat H_{MX}\hat P$ is the sum of two terms
\begin{eqnarray}\label{Hs}
 \hat{H} &=& \hat{H}_0+\hat V_C.
\end{eqnarray}
Here, the first term,
\begin{eqnarray}\label{H0}
 \hat{H}_0&=& \sum_{a\geq 1}|x_a\rangle\hbar\omega^x_a\langle x_a|
    + \sum_{k\geq 1} |xx_k\rangle\hbar\omega^{xx}_k\langle xx_k|,
\end{eqnarray}
describes non-interacting single- and bi-exciton states, and the second term
\begin{eqnarray}\label{HW}
 \hat{V}_C &=& \sum_{a\geq0}\sum_{k\geq 1}|x_a\rangle V^{x,xx}_{a,k}\langle xx_k|
 +h.c.,
\end{eqnarray}
represents the interband Coulomb interactions, $V^{x,xx}_{a,k}$, between the states, as well as the vacuum to bi-exciton couplings,  $V^{xx,x}_{k,0}$. Explicit representations for interaction matrix elements in terms of the single-particle couplings, and related matrix equations defining the single- and bi-exciton states are provided in Appendix~\ref{EffH}.

The dynamics restricted to the subspace of interest is fully defined by the projected propagator $\hat G(t)=\hat P \hat U(t) \hat P$ whose representation in the frequency domain is\cite{mukamel_book}
\begin{eqnarray}\label{PUP}
    \hat G(\omega) &=& i\left[\omega-\hat h_{eff}(\omega)+i\gamma\right]^{-1},
\end{eqnarray}
where $\gamma$ is the finite broadening associated with the exciton-phonon coupling. The non-local effective Hamiltonian entering this Green function can be partitioned into the sum of the diagonal and off-diagonal terms
\begin{eqnarray}\label{Heff}
    \hat h_{eff}(\omega) &=& \hat h(\omega) + \hat v(\omega),
\end{eqnarray}
which have the following forms
\begin{eqnarray}
   \label{h-intra}
    \hat h(\omega) &=& \hbar^{-1}\hat H_0 + \hat k_{d}(\omega),
    \\\label{w-inter}
    \hat v(\omega) &=& \hbar^{-1} \hat V_C + \hat k_{o}(\omega),
\end{eqnarray}
respectively. The first terms in Eqs.~(\ref{h-intra}) and (\ref{w-inter}) are components of the projected Hamiltonian $\hbar^{-1}\hat H$ (Eq.~(\ref{Hs})--(\ref{HW})) describing the propagation of the coupled single- and bi-exciton states. The second terms, accounting for the effect of the higher-multiplicity exciton states, are the diagonal, $\hat k_{d}(\omega)$, and off-diagonal, $\hat k_{o}(\omega)$, components of the non-local memory kernel, respectively.

The memory kernel components can be explicitly represented in the multi-exciton bases as
\begin{eqnarray}
   \label{K-intra}
    \hat k_{d}(\omega)&=&\hbar^{-2}\sum_{\bar n\bar m\geq3}\sum_{pq\geq1}
    		\left[
	\right.\\\nonumber &~&\left.	
		\sum_{ab\geq0} |x_a\rangle V^{x,\bar n}_{a,p}
    		\tilde G^{\bar n,\bar m}_{p,q}(\omega)V^{\bar m,x}_{q,b}\langle x_b|\right.
	\\\nonumber
        &+&\left. \sum_{kl\geq1}|xx_k\rangle V^{xx,\bar n}_{k,p}
    		\tilde G^{\bar n,\bar m}_{p,q}(\omega)V^{\bar m,xx}_{q, l}\langle xx_l|\right],
    \\\label{K-inter}
    \hat k_{o}(\omega)&=&\hbar^{-2}\sum_{\bar n\bar m\geq3}\sum_{pq\geq1}\sum_{a\geq0}\sum_{k\geq1}
    	\left[
	\right.\\\nonumber &~&\left.
	|x_a\rangle {V}^{x,\bar n}_{a,p}\tilde G_{p,q}^{\bar n,\bar m}(\omega)
	V^{\bar m,xx}_{q,l}\langle xx_l|
	\right.\\\nonumber
        &+&\left.|xx_l\rangle V^{xx,\bar m}_{l,q}
    		\tilde G_{q,p}^{\bar m,\bar n}(\omega)V^{\bar n,x}_{p, a}\langle x_a|\right],
\end{eqnarray}
where $V^{x,\bar n}_{a,p}$ ($V^{xx,\bar m}_{l,q}$) are the interaction matrix elements (Eq.~(\ref{HX12})) which couple single-exciton (bi-exciton) states with the states of multiplicity $\bar n\bar m\geq3$. The matrix elements $\tilde G^{\bar n,\bar m}_{p,q}(\omega)=\langle x^{\bar n}_p|\tilde G(\omega)|x^{\bar m}_q\rangle$ of the propagator
\begin{eqnarray}\label{QUQ}
\tilde G(\omega) = i\left[\omega - \hbar^{-1}\hat Q\hat H_{MX}\hat Q\right]^{-1},
\end{eqnarray}
describe the projected dynamics in the higher-multiplicity exciton space defined by $\hat Q$  (the lower right quadrant in Fig~\ref{Fig-Hmlt}).  

The use of projection operators allows us to map the propagator acting in the multi-exciton space (Eq.~(\ref{U-MX})) to the projected propagator acting in the space of single- and bi-exciton states (Eqs.~(\ref{PUP})--(\ref{QUQ})). This representation is exact, since the effect of the higher-multiplicity exciton states is fully accounted for through the memory kernel (Eqs.~(\ref{K-intra}) and (\ref{K-inter})). The dynamics of interest can now be interpreted as the uncoupled propagation within single- and bi-exciton manifolds described by the zeroth-order Green function,  
\begin{eqnarray}\label{g-oprt}
 \hat g(\omega)=i\left[\omega-\hat h(\omega)+i\gamma\right]^{-1},
\end{eqnarray}
and the scattering events between these manifolds induced by the interaction operator $\hat v(\omega)$.

\subsection{Single- and bi-exciton scattering model}
\label{esmd}

\begin{figure}[t]
\centerline{\includegraphics[width=3.0in]{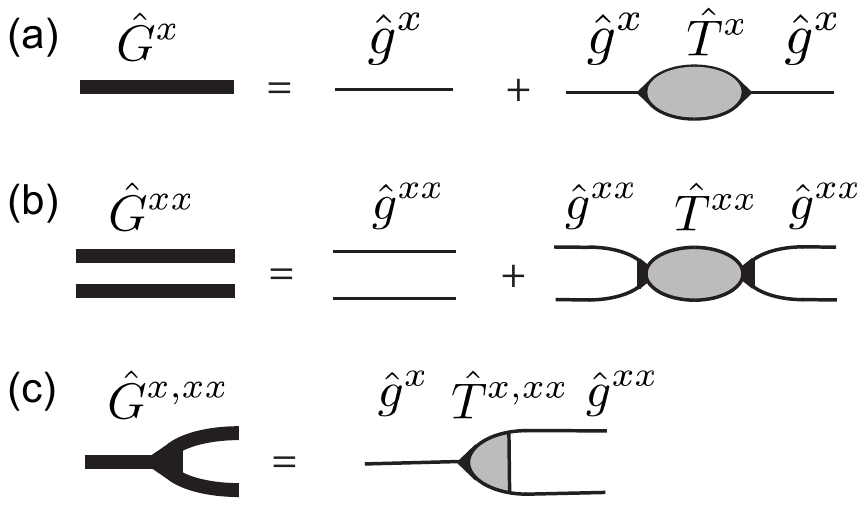}}
    \caption{Feynman diagram representation of the projected propagator, $\hat G$, in terms of the
    scattering operator, $\hat T$. Panels (a) and (b) show the single- and bi-exciton propagator components,
    respectively. (c) The interband component of the propagator mixing the single- and bi-exciton states.}
    \label{Fig-sdgr}
\end{figure}

To apply the scattering matrix formalism, we represent the projected Green function (Eq.~(\ref{PUP})) as a $2\times 2$ block matrix
\begin{eqnarray}\label{G-mtrx}
 \hat G(\omega) = \left( \begin{array}{cc}
					\hat G^{x}(\omega)    & \hat G^{x,xx}(\omega)\\
					\hat G^{xx,x}(\omega) & \hat G^{xx}(\omega)  \\
	\end{array} \right).
\end{eqnarray}
The Fourier transformation of Eq.~(\ref{G-mtrx})
\begin{eqnarray}
    \label{Gf-Four}
    \hat G(t)= \int_{-\infty}^\infty\frac{d\omega}{2\pi}\hat G(\omega)\exp{\left(-i\omega t\right)},
\end{eqnarray}
defines time-evolution of the single- and bi-exciton states
\begin{eqnarray}\label{X-Tevol}
&~&|x_a(t)\rangle = \sum_{b\geq1} G^{x}_{ab}(t)|x_b(0)\rangle
        +\sum_{k\geq1}G^{x,xx}_{a,k}(t) |xx_k(0)\rangle,\;\;\;\;\;\;
 \\\label{XX-Tevol}
&~&|xx_k(t)\rangle = \sum_{a\geq0}G^{xx,x}_{k,a}(t)|x_a(0)\rangle+
            \sum_{l\geq1}G^{xx}_{kl}(t) |xx_l(0)\rangle.
\end{eqnarray}
According to Eqs.~(\ref{X-Tevol}) and (\ref{XX-Tevol}), the matrix elements $G^{x}_{ab}(t)$ and $G^{xx}_{kl}(t)$ associated with the diagonal blocks in Eq.~(\ref{G-mtrx}) determine the intraband propagation, and the matrix elements $G^{x,xx}_{a,k}(t)$ associated with the off-diagonal blocks in Eq.~(\ref{G-mtrx}) describe the interband scattering processes mixing the single- and bi-exciton states.

Within the scattering matrix formalism, the propagator, $\hat G$, satisfies the following equation:\cite{economou_greenf}
\begin{eqnarray}\label{GT-rep}
    \hat G(\omega)= \hat g(\omega) + \hat g(\omega)\hat T(\omega)\hat g(\omega),
\end{eqnarray}
in which $\hat g(\omega)$ is the intraband zeroth-order Green function introduced in Eq.~(\ref{g-oprt}). In the modified block-matrix representation, this Green function is
\begin{eqnarray}\label{g-mtrx}
 \hat g(\omega) = \left( \begin{array}{cc}
					\hat g^{x}(\omega)    & 0\\
					0 & \hat g^{xx}(\omega)  \\
	\end{array} \right).
\end{eqnarray}
Here, the diagonal blocks, $\hat g^{x}(\omega)$ and $\hat g^{xx}(\omega)$, can be determined numerically using Eq.~(\ref{g-oprt}) with matrix inversion. Finally, the scattering operator in the same representation is
\begin{eqnarray}\label{T-mtrx}
 \hat T(\omega) = \left( \begin{array}{cc}
					\hat T^{x}(\omega)    & \hat T^{x,xx}(\omega)\\
					\hat T^{xx,x}(\omega) & \hat T^{xx}(\omega)  \\
	\end{array} \right),
\end{eqnarray}
containing both single-exciton (bi-exciton) component, $\hat T^x$ ($\hat T^{xx}$), and interband components, $T^{x,xx}$. The Feynman diagram representation of Eq.~(\ref{GT-rep}) is given in Fig.~\ref{Fig-sdgr}. To find the solution of Eqs.~(\ref{GT-rep})--(\ref{T-mtrx}), we need to know the form of the matrix elements of Eq.~(\ref{T-mtrx}).

\begin{figure}[t]
\centerline{\includegraphics[width=3.6in]{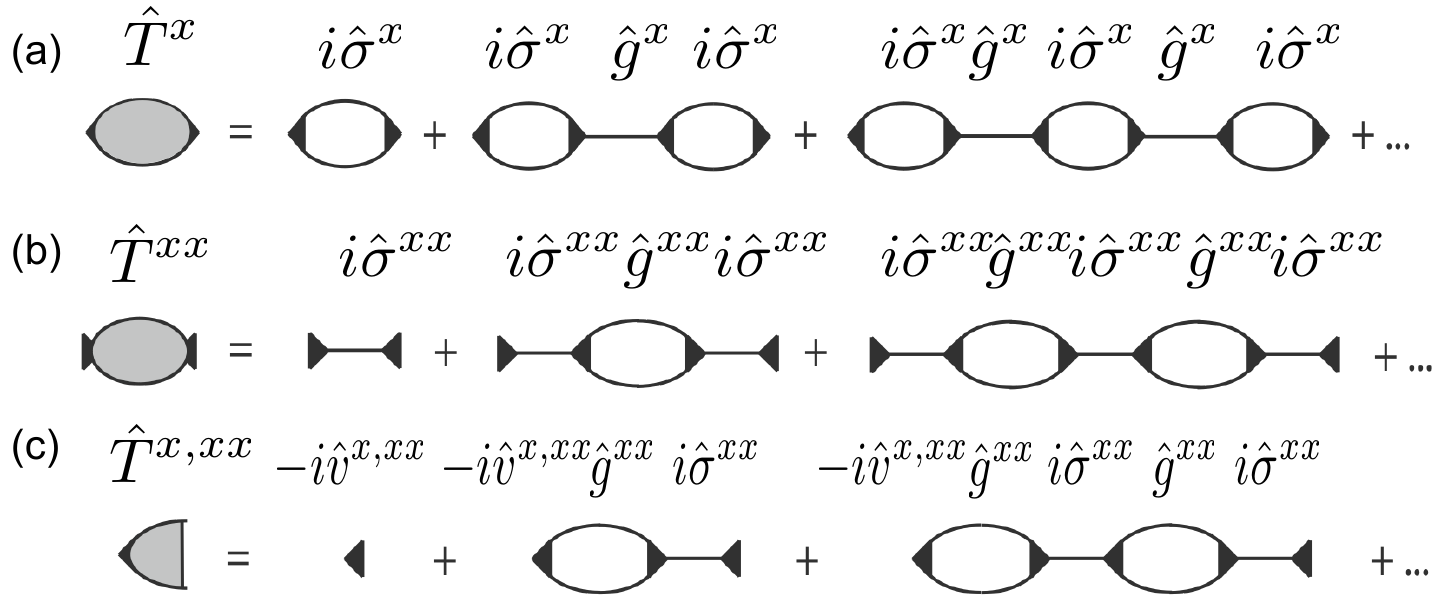}}
    \caption{Feynman diagram expansion for the scattering matrix, $\hat T$. Panels (a) and (b)
    describe even-order scattering events contributing to the single-exciton and bi-exciton scattering
    matrix respectively. (c) The odd-order scattering events changing the exciton multiplicity contribute
    to the interband scattering matrix. The latter processes give rise to CM.}
    \label{Fig-tdgr}
\end{figure}

To obtain a closed set of equations for the scattering matrix $\hat T$, the projected propagator (Eq.~(\ref{PUP})) should be expanded in a  power series of the interband coupling operator, $\hat v(\omega)$ (Eq.~(\ref{w-inter})). These expansion terms can be further regrouped to match the form of Eq.~(\ref{GT-rep}), leading to the diagrammatic expansion of the scattering operator shown in Fig.~\ref{Fig-tdgr}. 

In Fig.~\ref{Fig-tdgr}, panels~(a) and (b), represent the diagrams contributing to the diagonal scattering matrix blocks $\hat T^{x}(\omega)$ and $\hat T^{xx}(\omega)$, respectively. Each term there contains an $\it even$ number of vertices, reflecting the even number of interband scattering events. This leads to conservation of the excitons multiplicity, and to renormalization of their energies. The summation of this diagrammatic series in panels~(a) and (b) results in a set of linear equations for the single-exciton ($\bar n=x$) and the bi-exciton ($\bar n=xx$) scattering matrix elements:
\begin{eqnarray}
    \label{TF-intra}
    \sum_{kl}\left[\delta_{ik}\delta_{lj}-i\sigma^{\bar n}_{ik}(\omega)g^{\bar n}_{kl}(\omega)\right]
    T_{lj}^{\bar n}(\omega)= i\sigma^{\bar n}_{ij}(\omega).
\end{eqnarray}
Here, the self-energy matrix elements renormalize the bare single- and bi-exciton energies, and according to Fig.~\ref{Fig-tdgr}~(a) and (b) can be represented as
\begin{eqnarray}
    \label{Selfe-eq}
    \sigma_{ij}^{\bar n}(\omega)= i\hbar^{-2}\sum_{kl} v_{i,k}^{\bar n,\bar m}(\omega)
    g^{\bar m}_{kl}(\omega)v_{l,j}^{\bar m,\bar n}(\omega),
\end{eqnarray}
where $\bar m=xx$ ($\bar m=x$) if $\bar n=x$ ($\bar n=xx$).

According to Fig.~\ref{Fig-tdgr}~(c), the interband scattering matrix depends on $\hat T^{xx}$ and can be calculated from the following linear transformation: 
\begin{eqnarray}
  \label{TF-inter}
   T^{x,xx}_{a,l}(\omega)&=& -iv^{x,xx}_{a,l}(\omega)
   \\\nonumber&+&
   \sum_{mn} -iv^{x,xx}_{a,m}(\omega) g^{xx}_{mn}(\omega)T^{xx}_{nl}(\omega).
\end{eqnarray}
This scattering matrix accounts for the {\it odd}-order scattering events changing the multiplicity of the initial exciton state, and, therefore, describes the CM dynamics.

Equations~(\ref{GT-rep})--(\ref{TF-inter}), are exact, since they account for all terms entering the multi-exciton Hamiltonian (Eqs.~(\ref{HX})--(\ref{HX12})). Specifically, these terms determine the single- and bi-exciton binding energies and the interband interactions including the effects of the higher-multiplicity exciton states. In terms of the diagrammatic expansions shown in Figs.~\ref{Fig-sdgr} and \ref{Fig-tdgr}, the single and double lines associated with the components of $\hat g(\omega)$ and the vertices $\hat v(\omega)$ are dressed by these interactions. In practice, however, only approximate representations for $\hat g(\omega)$ and $\hat v(\omega)$ could be found. For instance, the multi-exciton binding energies can be determined approximately or even neglected. 

The more difficult task is the evaluation of the memory kernel entering the effective Hamiltonian (Eq.~(\ref{Heff})), since  the propagator $\tilde G(\omega)$ (Eq.~(\ref{QUQ})) cannot be calculated exactly. However, the kernel can be calculated approximately if the tri-exciton states are accounted for only. In this case the kernel will renormalize the single- and bi-exucotin resonances showing their hybridization with the tri-exciton ones.  If the tri-exciton (and higher-multiplicity exciton) poles are well-separated from the single- and bi-exciton resonances participating in the photoexcited dynamics, then the memory kernel can be dropped from the projected propagator, $\hat G(\omega)$ (Eq.~(\ref{PUP})). This situation is expected to take place in the vicinity of the AET, depending on the strength of the Coulomb couplings, $V^{x,\bar n}_{a,p}$ and $V^{xx,\bar m}_{n,q}$. 

At this point, we focus on the photoinduced dynamics in the vicinity of the AET only, and for the rest of the paper we assume that the memory kernel effect is negligible. Therefore, the scattering operator and the projected propagator can now be calculated by solving the set of linear Eqs.~(\ref{TF-intra})--(\ref{TF-inter}) where the zeroth-order propagator $\hat g(\omega)$ and interband interaction operator $\hat v(\omega)$ depend on the projected Hamiltonian (Eqs.~(\ref{Hs})--(\ref{HW})) only and Eqs.~(\ref{PUP}), (\ref{G-mtrx}), and (\ref{Gf-Four})), respectively. 

\section{Photoinduced dynamics in Liouville space}
\label{prdl}

In this section, we consider the carrier dynamics in an ensemble of NCs excited by a pump pulse whose fluence is adjusted so that no more than a single photon is absorbed per NC. This results in the preparation of no more than one single- or bi-exciton state in each NC interacting with photons leading to a total population produced by the pulse which can be determined by the ensemble average. 

\begin{figure}[t]
\centering
{\includegraphics[width=3.2in]{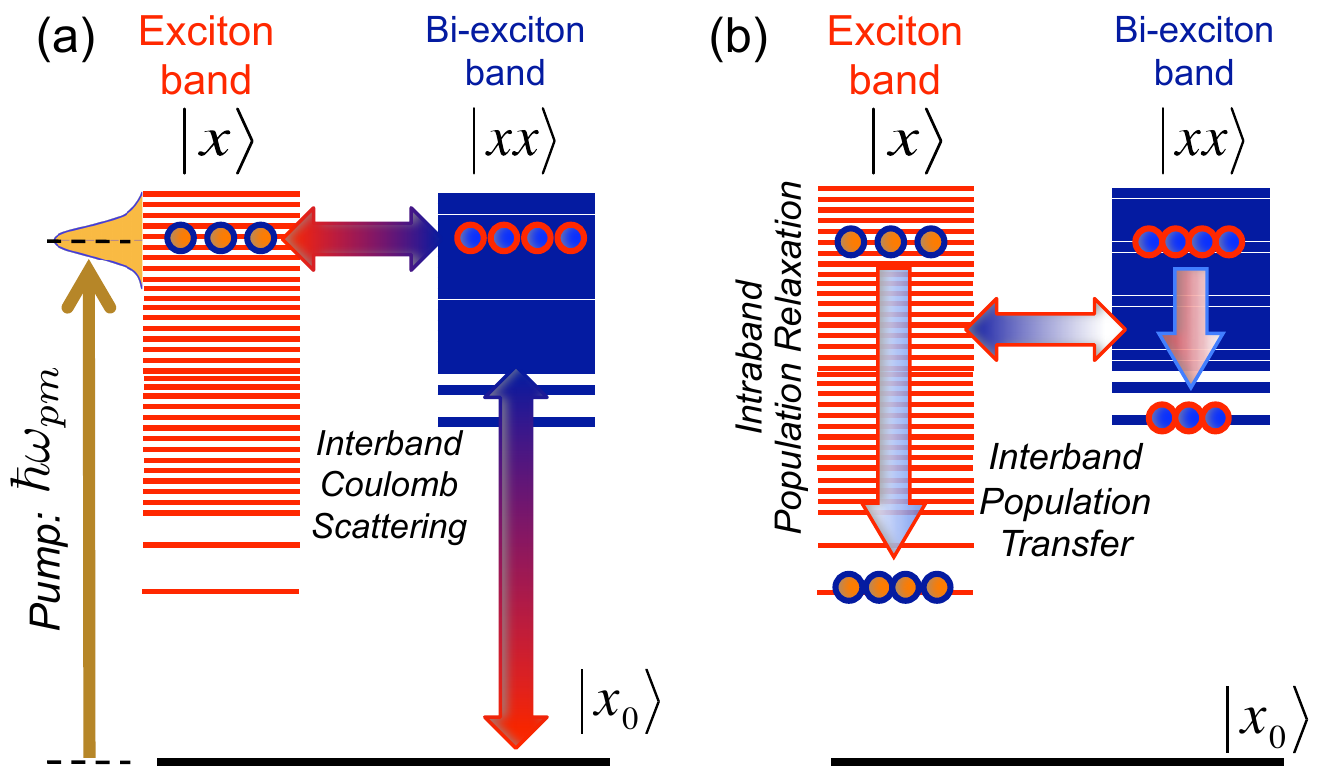}}
\caption{Level diagram of CM dynamics in ensemble of NCs.
	(a) Photoexcitation by a pump pulse with central frequency, $\omega_{pm}$, and finite spectral widths results in the 
	generation of single- and bi-exciton populations in which the Coulomb scattering mixes all interband and intraband 
	dipole transitions present in Eq.~(\ref{H-opt}). (b) During the population relaxation, both the intraband and the interband 
	processes are mixture of the phonon-assisted processes and Coulomb scattering events.}
\label{Fig-pdyn}
\end{figure}

The photoinduced ensemble dynamics is illustrated in Fig.~\ref{Fig-pdyn}: Panel (a) shows the exciton photogeneration event which occurs on the pump timescale ranging between $50-100$~fs. During the photogeneration, the relative number of single- and bi-exciton states produced by the pump is determined by the interband scattering processes. The photogenerated populations further relax on the timescale of $1-10$~ps as shown in panel (b). As we demonstrate below, this relaxation includes phonon-assisted cooling to the bottom of the single- and bi-exciton bands {\em mixed} with the interband population transfer due to the Coulomb scattering. The population from the bottom of the bi-exciton band finally decays to the lowest single-exciton states through Auger recombination within $\gtrsim 10$~ps. This process (not shown in Fig.~\ref{Fig-pdyn}) is typically employed for the experimental determination of the bi-exciton production yield, and has no contribution to QE. Therefore, we do not consider this process in this paper.

To include the interaction with the optical pump, we extend the projected Hamiltonian as
\begin{equation}\label{H'}
	\hat{H}_{opt} = \hat{H}+\hat{V}(t)
\end{equation}
where the following time-dependent term is added
\begin{eqnarray}\nonumber
 \hat{V}(t) &=& - E(t)\left[\sum_{ab\geq0}|x_a\rangle \mu^{x}_{ab}\langle x_b|
 			+\sum_{kl\geq1}|xx_k\rangle\mu^{xx}_{kl}\langle xx_l|\right]
 \\\label{H-opt}
    			&-&  E(t)\sum_{ak\geq1}\left(|x_a\rangle \mu^{x,xx}_{ak}\langle xx_k|  			
			+|xx_k\rangle \mu^{xx,x}_{ka}\langle x_a|\right).\;\;\;\;\;\;\;\;
\end{eqnarray}
Here, the optical pulse 
\begin{equation}\label{pp}
    E(t)={\cal E}_{pm}(t)\exp(-i\omega t)+c.c.
\end{equation}
is characterized by the absolute value of the envelope function, ${\cal E}_{pm}(t)$, with the widths, $\tau_{pm}$, describing the pulse duration, and the central frequency, $\omega_{pm}$. \cite{ftn-01}{The pump envelope and spatial phases do not contribute to the population dynamics and therefore are dropped.} Details of the derivation of $\hat{H}_{opt}$ are given in Appendix~\ref{eofh}.

According to Eq.~(\ref{H-opt}), the optical field interacts with all possible transition dipoles which couple the single-exciton states $\mu^{x}_{ab}$, the bi-exciton states $\mu^{xx}_{kl}$, and the single- to bi-exciton states $\mu^{x,xx}_{ak}={\mu^{xx,x}_{ka}}^*$. No permanent dipoles are present in this ensemble, i.e. $\mu^{x}_{aa}=\mu^{xx}_{kk}=0$. In general, all these transitions are allowed due to the Coulomb scattering processes.

Next, we employ the density matrix formalism to include the dissipation processes due to the coupled phonon bath. Within this formalism, the dynamics of interest are fully described by the Liouville equation:
\begin{equation}\label{LiouEq}
    \dot{\hat\rho}(t) = (i\hbar)^{-1}\left[\hat H_{opt},\hat\rho(t)\right]
    +\hat{\cal R}\hat\rho,
\end{equation}
where the time-dependent density operator is a $2\times 2$ block matrix
\begin{equation}\label{rho-mtrx}
	\hat\rho(t) = \left( \begin{array}{cc}
					\hat \rho^{x}(t)    & \hat \rho^{x,xx}(t)\\
					\hat \rho^{xx,x}(t) & \hat \rho^{xx}(t)  \\
	\end{array} \right),
\end{equation}
containing single-exciton $\hat \rho^{x}(t)$ and bi-exciton $\hat \rho^{xx}(t)$ components, and coherences between single- and bi-exciton states $\hat \rho^{xx,x}(t)$. The specific form of the relaxation term, ${\hat{\cal R}\hat\rho}$, in Eq.~(\ref{LiouEq}) depends on the specific exciton-phonon interaction model.

\subsection{Phonon-assisted relaxation model}
\label{psrm}

To describe the phonon-assisted dynamics, we assume that the phonon bath has a continuous spectral density, and there is no phonon bottleneck.\cite{guyotsionnest05,pandey08,kilina07,kilina09} An explicit form of the spectral density depends on the environment model with adjustable parameters such as spectral widths and electron-phonon coupling strengths. The simplest model which can be employed in our case is the model of single- and bi-exciton states linearly coupled to the phonon coordinates $\{q_\alpha\}_{\alpha=1,2,3,\dots}$. The related Hamiltonian is
\begin{eqnarray}\label{Hxp-full}
 \hat{H}_{ep} &=& \hat H_{i} + H_{p},
\end{eqnarray}
where the exciton-phonon interaction term
\begin{eqnarray}\nonumber
 \hat{H}_{i} &=& \sum_{ab;\alpha}|x_a\rangle Y^{x}_{ab;\alpha}q_\alpha\langle x_b|
 +\sum_{kl;\alpha}|xx_k\rangle Y^{xx}_{kl;\alpha}q_\alpha\langle xx_l|\;\;\;\;
 \\\label{Hxp-int} &+&
 	\sum_{ak;\alpha}\left(|x_a\rangle Y^{x,xx}_{ak;\alpha}q_\alpha\langle xx_k|
   +|xx_k\rangle Y^{xx,x}_{ka;\alpha}\langle x_a|\right),
\end{eqnarray}
contains the intraband single-exciton (bi-exciton) coupling matrix elements $Y^{x}_{ab;\alpha}$ ($Y^{xx}_{ab;\alpha}$) to $\alpha$-th phonon mode, and the interband coupling matrix elements $Y^{xx,x}_{ka;\alpha}$. The connections between the former quantities and the {\em electron}-phonon coupling constants from the many-body Hamiltonian are given in Appendix~\ref{RxxH}. The second term in Eq.~(\ref{Hxp-full}) is the uncoupled phonon Hamiltonian whose form depends on the specific environment model.

Assuming weak exciton-phonon coupling, we follow a standard projection operator method to eliminate the bath degrees of freedom  resulting in the Markov approximation for ${\hat{\cal R}}\hat{\rho}$.\cite{kubo95,vankampen92} The basis set in which the equilibrium  (Gibbs) distribution, $\bar\rho$, can be recovered as the zero eigenfunction of the relaxation operator, i.e. ${\hat{\cal R}}\bar{\rho}=0$, is the {\em quasiparticle} basis $\{|\bar\xi\rangle\}_{\bar\xi=0,1,2,\dots}$ formed by the eigenstates of the total projected Hamiltonian (Eqs.~(\ref{Hs})--(\ref{HW})).\cite{dahlbom00} Therefore, we consider the population relaxation dynamics in this preferred basis. 

After introducing the quasiparticle energies, $\hbar\omega_{\bar\xi}$, and further using the interaction representation for the density operator, i.e. $\tilde\rho_{\bar\xi^{'}\bar\zeta^{'}}(t)=e^{-i\omega_{\bar\xi\bar\zeta}t}\rho_{\bar\xi\bar\zeta}(t)$ with $\omega_{\bar\xi\bar\zeta}=\omega_{\bar\xi}-\omega_{\bar\zeta}$, we recast the Liouville Eq.~(\ref{LiouEq}) in the absence of the optical pulse ($\hat V(t)=0$)   in the quasiparticle basis. This results in the Redfield Equation:
\begin{eqnarray}\label{Redfeq}
   \dot{\tilde \rho}_{\bar\xi\bar\zeta}(t) &=&
   \sum_{\bar\xi^{'}\bar\zeta^{'}}e^{-i(\omega_{\bar\xi\bar\zeta}-\omega_{\bar\xi^{'}\bar\zeta^{'}})t}
    {\cal R}_{\bar\xi\bar\zeta;\bar\xi^{'}\bar\zeta^{'}}
    \tilde\rho_{\bar\xi^{'}\bar\zeta^{'}}(t),
\end{eqnarray}
where ${\cal R}_{\bar\xi\bar\zeta;\bar\xi^{'}\bar\zeta^{'}}$ is the relaxation tensor.\cite{ftn-13}

The interaction representation allows us to apply the so-called secular approximation, eliminating the rapidly oscillating terms containing $\omega_{\bar\xi\bar\zeta}-\omega_{\bar\xi^{'}\bar\zeta^{'}}\neq 0$. Respectively, the remaining Redfield tensor components,\cite{ernst90}
\begin{eqnarray}\label{Redfield-Ts-P}
    R_{\bar\xi\bar\xi;\bar\xi^{'}\bar\xi^{'}}&=&
    	- \delta_{\bar\xi\bar\xi^{'}}\sum_{\bar\sigma\neq\bar\xi}\Gamma_{\bar\sigma\bar\xi}
	+\Gamma_{\bar\xi\bar\xi^{'}},
	\\\label{Redfield-Ts-C}
	R_{\bar\xi\bar\zeta;\bar\xi\bar\zeta}&=&- \frac{1}{2}\sum_{\bar\sigma\neq\bar\xi}\Gamma_{\bar\sigma\bar\xi}
	-\frac{1}{2}\sum_{\bar\sigma\neq\bar\zeta}\Gamma_{\bar\sigma\bar\zeta}-\gamma_{\bar\xi\bar\zeta},
\end{eqnarray}
correspond to uncoupled equations for the population relaxation and coherence dephasing. Eqs.~(\ref{Redfield-Ts-P}) and (\ref{Redfield-Ts-C}) contain the population relaxation and pure dephasing rates:
\begin{eqnarray}\label{prlx-rate}
    \Gamma_{\bar\xi\bar\xi^{'}} &=& \frac{1}{\hbar^2}\sum_{\alpha\alpha^{'}}
        Y_{\bar\xi\bar\xi^{'};\alpha}Y_{\bar\xi^{'}\bar\xi;\alpha^{'}}
        C_{\alpha\alpha^{'}}(\omega_{\bar\xi\bar\xi^{'}}),
   \\\label{dphas-rate}
    \gamma_{\bar\xi\bar\zeta} &=&\frac{1}{\hbar^2}\sum_{\alpha\alpha^{'}}
    \left(Y_{\bar\xi\bar\xi;\alpha}-Y_{\bar\zeta\bar\zeta;\alpha}\right)
    \\\nonumber&\times&
    \left(Y_{\bar\xi\bar\xi;\alpha^{'}}-Y_{\bar\zeta\bar\zeta;\alpha^{'}}\right)
        \tilde C^{'}_{\alpha\alpha^{'}}(0),
\end{eqnarray}
respectively. Here, $\tilde C_{\alpha\alpha^{'}}(\omega)$ denotes the Fourier transform of the phonon correlation function, $C_{\alpha\alpha^{'}}(\tau)=\langle e^{\hat H_{p}\tau} \hat q_\alpha e^{-i\hat H_{p}\tau}\hat q_{\alpha'}\rangle_{eq}$, and it has both real $C_{\alpha\alpha^{'}}^{'}(\omega)$ and imaginary $C_{\alpha\alpha^{'}}^{''}(\omega)$ parts. The explicit representation for the correlation function depends on the chosen form of $H_p$, i.e. on the specific relaxation model. 

The products of the off-diagonal quasiparticle-phonon coupling constants entering the population relaxation rate (Eq.~(\ref{prlx-rate})) can be expressed in terms of the exciton-phonon matrix elements entering Eq.~(\ref{Hxp-int}) as:
\begin{eqnarray}\label{Yqp-o}
    Y_{\bar\xi\bar\xi^{'};\alpha}Y_{\bar\xi^{'}\bar\xi;\alpha^{'}} = 
    \sum_{ll'rr'}\bar\Lambda_{lr}(\omega_{\bar\xi}) \bar\Lambda_{l'r'}(\omega_{\bar\xi'})
    Y_{ll';\alpha}Y_{r'r;\alpha^{'}}.
\end{eqnarray}
Finally, the matrix element of the diagonal quasiparticle-phonon coupling determining the pure dephasing rate (Eq.~(\ref{prlx-rate})) is
\begin{eqnarray}\label{Yqp-d}
    Y_{\bar\xi\bar\xi;\alpha}=\sum_{lr}\bar\Lambda_{lr}(\omega_{\bar\xi})Y_{lr;\alpha}.
\end{eqnarray}
In  Eqs.~(\ref{Yqp-o}) and (\ref{Yqp-d}), $\bar\Lambda_{lr}(\omega_{\bar\xi})=\lim_{\gamma\rightarrow 0}{\rm res}\{G_{lr}(\tilde\omega_{\bar\xi})\}$ is the transition amplitude given by the Green function residue in the limit of infinitesimal imaginary part, $\gamma$, of the poles.\cite{ftn-02} Here and below, we use the convention that the summation indices for single- and bi-exciton states (particularly those in Eqs.~(\ref{Yqp-o}) and (\ref{Yqp-d})) run over all single- and bi-exciton states, unless the superscripts $x$ or $xx$ constraining their range are used (e.g. in Eq.~(\ref{Hxp-int}) for $Y^{x}_{ab;\alpha}$, $Y^{xx}_{kl;\alpha}$, and $Y^{x,xx}_{ak;\alpha}$).

The dependence of the population relaxation rate (Eqs.~(\ref{prlx-rate}) and (\ref{Yqp-o})) on the transition amplitude indicates that these  exciton scattering processes are involved in phonon-assisted cooling. Some of them, as we demonstrated in Sec.~\ref{esmd}, change the multiplicity of the initial states, and, therefore, can be considered as generalized impact ionization and Auger recombination processes. Accordingly, we argue that the phonon-assisted cooling and impact ionization dynamics giving rise to CM are generally coupled. However, we demonstrate in Sec.~\ref{wclm}, that the {\em intraband} phonon-assisted cooling and the interband impact ionization and Auger recombination processes can be decoupled in the limit of weak Coulomb coupling.

By applying the secular approximation, we significantly simplify the description of the phonon-assisted dynamics. However,  the validity of our approximation for NCs is based on the following delicate interplay between the number of quantum states and their energy separations: In the region of high DOS, some closely lying levels may have $\omega_{\bar\xi\bar\zeta}-\omega_{\bar\xi^{'}\bar\zeta^{'}}\approx 0$,  potentially leading to the breakdown of the secular approximation. On the other hand, we expect that, due to the same high DOS, there are enough terms in the sum of Eq.~(\ref{Redfeq}) containing these slowly-oscillating phases to cancel out their contributions. Therefore, the chosen secular approximation must be validated using numerical simulations, for specific materials.

\subsection{Photoexcited population dynamics and QE}
\label{pmpr}

The central quantity describing CM is QE which can be calculated as 
\begin{eqnarray}
\label{QE}
    QE =\frac{2N_{xx}(\tau)+N_x(\tau)}{N_{xx}(\tau)+N_x(\tau)},
\end{eqnarray}
where $ N_x(\tau)=tr\hat\rho^x(\tau)$ and $N_{xx}(\tau)=tr\hat\rho^{xx}(\tau)$ are the total non-equilibrium single-exciton and bi-exciton populations, respectively. Their dependence on the delay time, $\tau$, measured from the center of the pump pulse, allows one to calculate both the QE due to the photogeneration event and the total QE after the population relaxation. The latter is typically measured in optical experiments.

\begin{figure}[t]
\centering
{\includegraphics[width=3.5in]{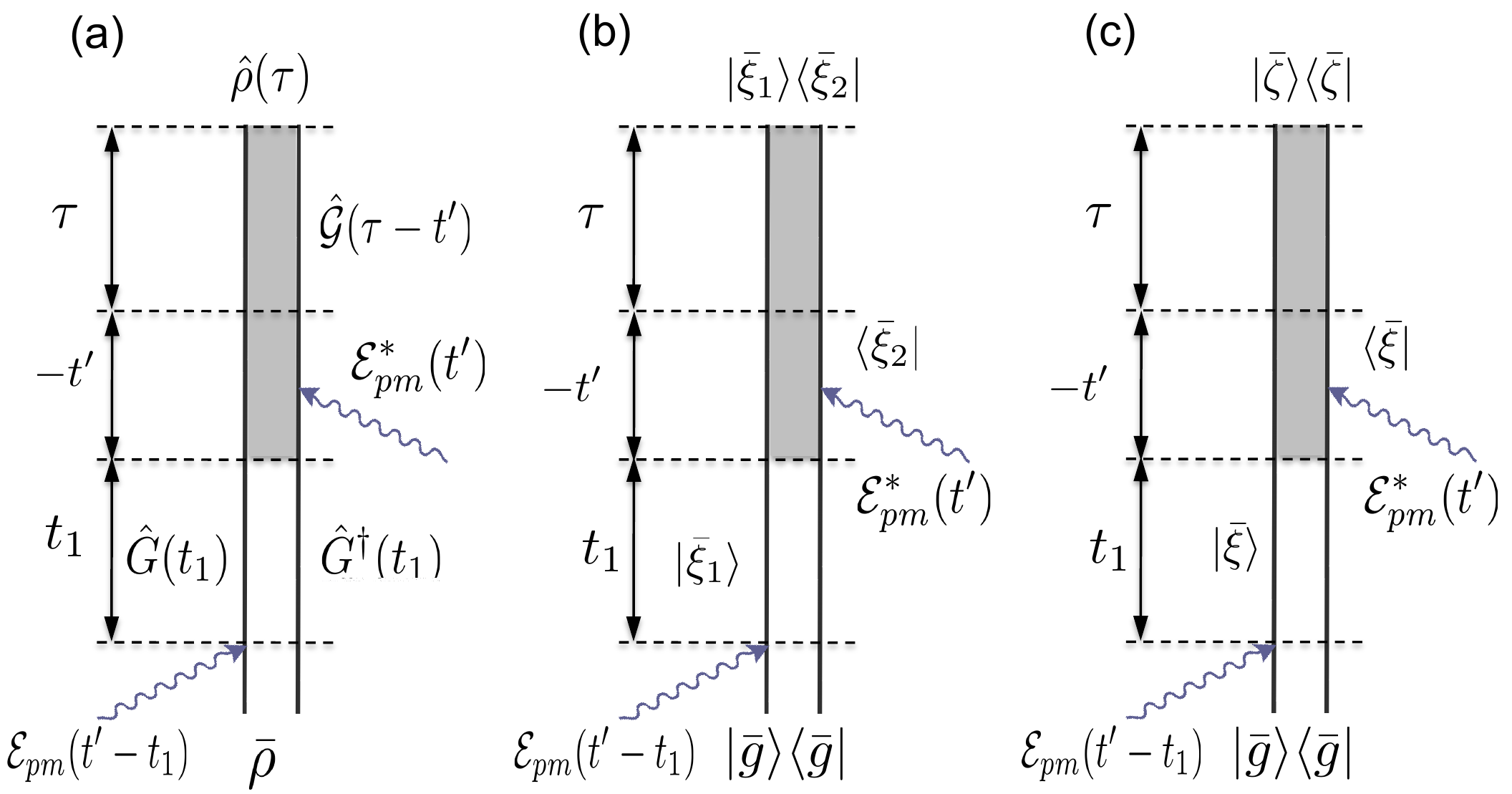}}
\caption{Double-sided Feynman diagram representation of the nonequilibrium density operator, $\hat\rho(\tau)$, prepared by the pump pulse and propagated during delay time, $\tau$. The times $t_1$ and $t'$, are the integration variables, and $\bar\rho$ is the equilibrium density operator. $\hat\rho(\tau)$ can be partitioned into two components reflecting the contributions of (b) the {\it quasiparticle} coherences and (c) the {\it quasiparticle} populations.}
\label{Fig-dsdg}
\end{figure}

The calculation of QE requires the solution of the Liouville equation given by Eq.~(\ref{LiouEq}) for single- and bi-exciton populations, $\rho_{s}$. The solution of Eq.~(\ref{LiouEq}) can be obtained by using the fact that the coupling between the optical field and the exciton states is weak compared to the transition energies. This results in the second order perturbation expression associated with the double-sided Feynman diagram presented in Fig.~\ref{Fig-dsdg}~(a).\cite{mukamel_book} The resulting population matrix element is 
\begin{eqnarray}
\label{NGF}
    \rho_{s}(\tau)&=&-2\hbar^{-2}{\rm Re}
    \sum_{l_0l_1l_2}\sum_{r_0r_1r_2}
    \\\nonumber&~&
    \mu_{l_1l_0}\bar\rho_{l_0r_0}\mu_{r_1r_2}
    \\\nonumber&\times&
    \int_{-\infty}^\infty dt^{'}\int_{0}^\infty dt_1
    {~\cal G}_{ss,l_2r_2}(\tau-t^{'})
    \\\nonumber&\times&
    G_{l_2 l_1}(t_1)G^*_{r_1 r_0}(t_1)
    \\\nonumber&\times&
    {\cal E}_{pm}(t^{'}){\cal E}_{pm}(t^{'}-t_1)
    e^{i\omega_{pm} t_1},
\end{eqnarray}
where $G_{lr}(t)$ is the matrix element of the projected propagator (Sec.~\ref{rdyn}), and  ${\cal G}_{ll,l_3r_1}(\tau-t^{'})$ is the matrix element of the Redfield equation (Eq.~(\ref{Redfeq})) Green function transformed to the bare single- and bi-exciton basis. Finally, $\bar\rho_{lr}$ is the matrix element of the equilibrium density operator,\cite{ftn-03}
\begin{eqnarray}
\label{rho-eq}
    {\bar\rho}&=& |x_0\rangle\langle x_0|
    \\\nonumber
    &+&\sum_{k\geq 1}\left(|x_0\rangle\bar\Lambda^{x,xx}_{0,k} \langle xx_k|
    - |xx_k\rangle\bar\Lambda^{xx,x}_{k,0} \langle x_0|\right)
 \\\nonumber &+&
    \sum_{a\geq 1}\left(|x_0\rangle \bar\Lambda^{x}_{0a} \langle x_a|
    +|x_a\rangle \bar\Lambda^{x}_{a0} \langle x_0|\right).
\end{eqnarray}
with
\begin{eqnarray}
\label{LmbdCM-eq}
   &~&\bar\Lambda^{x,xx}_{0,k}=-\bar\Lambda^{xx,x}_{k,0}= -\frac{V^{x,xx}_{0,k}}{\hbar\omega_k},
\\\nonumber\label{LmbdX-eq}
    &~&\bar\Lambda^{x}_{0a}=\bar\Lambda^{x}_{a0}=\sum_{k\geq 1}\frac{V^{x,xx}_{0,k}V^{xx,x}_{k,a}}
    	{\hbar^2\omega^x_a\omega^{xx}_{k}}.
\end{eqnarray}

Equation~(\ref{NGF}) can be used for numerical calculations of the QE. This expression is quite general. Its form does not assume that the secular approximation (allowing the decoupling of the coherence and population relaxation dynamics) is used. For further analysis, we partition the contributions to $\rho_s(\tau)$ induced by the optical excitation of the quasiparticle populations and coherences. For this purpose, we represent the time-dependent matrix elements of the projected propagator as 
\begin{eqnarray}
\label{G-xres}
    G_{lr}(t) = \sum_{\bar\xi}\Lambda_{lr}(\omega_{\bar\xi})e^{-i\tilde\omega_{\bar\xi} t},
\end{eqnarray}
where the quasiparticle complex frequencies $\tilde\omega_{\bar\xi} = \omega_{\bar\xi}-i\gamma_{\bar\xi}$ are the poles of $G_{lr}(\omega)$, and $\Lambda_{lr}(\omega_{\bar\xi})={\rm res}\{G_{lr}(\tilde\omega_{\bar\xi})\}$ are the complex transition amplitudes given by the Green function residue.\cite{ftn-04} Eq.~(\ref{G-xres}) clarifies the physical meaning of the latter quantity showing that this is a probability amplitude for the transition between $l$ and $r$ states in the single- and bi-exciton basis associated with the propagation of the quasiparticle state, $|\bar\xi\rangle$.

Substitution of Eq.~(\ref{G-xres}) into Eq.~(\ref{NGF}) and partitioning the quasiparticle coherence and population dynamics (secular approximation) allows us to recast $\rho_s$ into a sum of the two terms,
\begin{eqnarray}
\label{rho-cp}
    \rho_{s}(\tau)&=& c_{s}(\tau)+n_{s}(\tau).
\end{eqnarray}
Here, the first term corresponds to the double-sided diagram shown in Fig.~\ref{Fig-dsdg} (b). It describes the contributions of the quasiparticle coherences: 
\begin{eqnarray}
\label{rhoc-qprt}
    c_{s}(\tau)&=&
   \sum_{\bar\xi_1\bar\xi_2}\sum_{s_0}
    \mu_{ss_0}(\bar\xi_1)\mu_{ss_0}^*(\bar\xi_2)
    e^{-i\tilde\omega_{\bar\xi_1\bar\xi_2}\tau}
\\\nonumber&\times&
   {\cal I}(\tilde\omega_{\bar\xi_10}-\omega_{pm};\tilde\omega_{\bar\xi_20}-\omega_{pm}),
\end{eqnarray}
where $\mu_{ss_0}(\bar\xi_1)$ is the projection of the transition dipole moment between the quasiparticle ground and $\bar\xi$-th states onto single-/bi-exciton states,
\begin{eqnarray}
\label{mu-eigs}
   \mu_{ss_0}(\bar\xi) &=&
    \sum_{l_1l_2}\Lambda_{sl_1}(\omega_{\bar\xi})\mu_{l_1l_2}\bar\rho_{l_2 s_0},
\end{eqnarray}
containing the matrix elements, $\bar\rho_{l_2 s_0}$, of the equilibrium density operator (Eqs.~(\ref{LmbdCM-eq})).\cite{ftn-05} Note that $\mu_{ss_0}(\bar\xi)$ mixes the interband and the intraband dipole transitions entering the optical interaction term of the Hamiltonian (Eq.~(\ref{H-opt})), and determine all possible photogeneration pathways.   

The pulse self-convolution function in Eq.~(\ref{rhoc-qprt}) is
\begin{eqnarray}
\label{Ic-eigs}\nonumber
    {\cal I}(\tilde\omega_{\bar\xi_10}&-&\omega_{pm};\tilde\omega_{\bar\xi_20}-\omega_{pm})=
    \frac{1}{\hbar^2}\int_{-\infty}^\infty dt^{'}\int_{0}^\infty dt_1
    \\&\times&
    \theta(\tau-t') e^{i\tilde\omega_{\bar\xi_{1}\bar\xi_2}t^{'}}
   \left[e^{-i\left(\tilde\omega_{\bar\xi_10}-\omega_{pm}\right)t_1}
   \right.\\\nonumber&+&\left.
   e^{i\left(\tilde\omega^*_{\bar\xi_20}-\omega_{pm}\right)t_1}\right]
   {\cal E}_{pm}(t^{'}) {\cal E}_{pm}(t^{'}-t_1).
   \end{eqnarray}
This function is weighted by the coherence between the quasiparticle excited and ground states $\tilde\omega_{\bar\xi_i,0}=\tilde\omega_{\bar\xi_i}-\tilde\omega_{0}-\gamma_{\bar\xi_i 0}$, and by the excited state coherences characterized by  $\tilde\omega_{\bar\xi_i,\bar\xi_j}=\tilde\omega_{\bar\xi_i}-\tilde\omega_{\bar\xi_j}-\gamma_{\bar\xi_i \bar\xi_j}$, where the dephasing rates, $\gamma_{\bar\xi_i g}$ and $\gamma_{\bar\xi_i \bar\xi_j}$, are determined by Eq.~(\ref{dphas-rate}).

The second term in Eq.~(\ref{rho-cp}), represented by the double-sided diagram in Fig.~\ref{Fig-dsdg} (c), describes the contributions of the quasiparticle populations 
\begin{eqnarray}
\label{rhop-qprt}
    n_{s}(\tau)&=&
   \sum_{lr} \sum_{\bar\xi\bar\zeta}
	\left[ \bar\Lambda_{ss}(\omega_{\bar\zeta})\bar{\cal G}_{\bar\zeta,\bar\xi}(\tau)
	\bar\Lambda_{lr}(\omega_{\bar\xi})\right]
\\\nonumber&~&
     \mu_{ls_0}(\bar\xi)\mu^*_{rs_0}(\bar\xi)
   {\cal I}(\tilde\omega_{\bar\xi 0}-\omega_{pr}).
\end{eqnarray}
Here, $\bar\Lambda_{lr}(\omega_{\bar\xi})=\lim_{\gamma\rightarrow 0}\Lambda_{lr}(\omega_{\bar\xi})$, and $\bar{\cal G}_{\bar\zeta,\bar\xi}(\tau)\equiv{\cal G}_{\bar\zeta\bar\zeta,\bar\xi\bar\xi}(\tau)$ denotes the quasiparticle population relaxation component of the Green function associated with the Redfield Equation. This Green function can be found in the standard way by using the eigenstates and eigenvalues of the Redfieled  operator (Eq.~(\ref{Redfield-Ts-P})).\cite{Zwanzig_NSM,mukamel_book,dahlbom00} If the high DOS does not allow the diagonalization of the relaxation operator, then $n_{s}(0)$ should be considered as the initial condition for the numerical solution of the Redfield equation (Eq.~(\ref{Redfeq})).\cite{ftn-06} Finally, the pulse self-convolution function in Eq.~(\ref{rhop-qprt}) simplifies to the form:
\begin{eqnarray}
\label{Ip-eigs}
    {\cal I}(\tilde\omega_{\bar\xi 0}&-&\omega_{pm})=
    \frac{2}{\hbar^2}\int_{-\infty}^\infty dt^{'}\int_{0}^\infty dt_1
    e^{-\gamma_{\bar\xi g}t_1}
    \\\nonumber&\times&
    \cos\left[\left(\omega_{\bar\xi 0}-\omega_{pm}\right)t_1\right]
   {\cal E}_{pm}(t^{'}) {\cal E}_{pm}(t^{'}-t_1),
   \end{eqnarray}
where we neglect the population relaxation processes during the interaction with the pulse. 

The representation given by Eqs.~(\ref{rho-cp})--(\ref{Ip-eigs}), provides a connection with the sum-over-eigenstates representation shown in Fig.~\ref{Fig-dsdg} (b) and (c). In this representation, there is an additional Liouville space pathway contribution to the single- and bi-exciton populations associated with the propagation of the ground state wave packet.\cite{mukamel_book} This term can, in principle, contribute to the CM dynamics if Coulomb coupling between the vacuum and bi-exciton states is strong enough to make the lowest excited state energy comparable with the thermal energy, $k_BT$, i.e. $V^{xx,0}\sim 2E_g-k_BT$. Since, the latter condition is not satisfied in NCs where typically $E_g\gg V^{xx,0}\gg k_BT$, we do not consider this pathway.

Finally, one can expect that the contribution of the quasiparticle coherences (Eq.~(\ref{rhoc-qprt})) to QE can become negligible compared to the quasiparticle populations (Eq.~(\ref{rhop-qprt})). This could happen, since spectral widths of ultrafast pump pulse can excite a significantly large number of states (Fig.~\ref{Fig-pdyn}(a)), whose phases entering Eq.~(\ref{rhoc-qprt}) through Eq.~(\ref{Ic-eigs}) add destructively. This assumption can be checked for specific materials through numerical evaluation of the related terms. 

\section{Limit of weak Coulomb coupling}
\label{wclm}

In this section, we consider the Exciton Scattering Model, developed in Secs.~\ref{xdhs} and \ref{prdl}, in the limiting case of weak Coulomb coupling. This limit is important for applications and assumes that the Coulomb matrix elements between single- and bi-exciton states are much smaller than the energy differences between these levels and/or  much smaller than the level broadening, i.e. $V_{a,k}^{x,xx}\ll\hbar~(|\omega^x_a-\omega^{xx}_k|,\gamma^{x,xx}_{a,k})$. As we demonstrate below, both carrier photogeneration and population relaxation dynamics can be described using no higher than second-order processes in the Coulomb  expansion. For this purpose, we use Eqs.~(\ref{rho-cp})--(\ref{Ip-eigs}), with the Green function components calculated in this limit.

\subsection{Time-domain Green function}

To find the Green functions,  we, first, represent the single-exciton ($\bar n = x$) and bi-exciton ($\bar n = xx$) free propagators  (Eq.~(\ref{g-oprt})) as
\begin{eqnarray}\label{g-wc}
    g^{\bar n}_{kl}(\omega) =\frac{i\delta_{kl}}{\left(\omega-\tilde\omega^{\bar n}_k\right)},
\end{eqnarray}
where the complex frequency, $\tilde\omega^{\bar n}_{k}=\omega^{\bar n}_k-i\gamma^{\bar n}_k$, contains the $k$-th frequency, $\omega^{\bar n}_k$, from the projected Hamiltonian (Eq.~(\ref{H0})), and the related dephasing rate, $\gamma^{\bar n}_k$.

If the interband Coulomb interaction is weak, the CM dynamics becomes dominated by the Born interband scatting represented by the first vertex diagram in Fig.~\ref{Fig-tdgr}~(c). According to Eq.~(\ref{TF-inter}), the scattering matrix elements in the Born approximation become
\begin{eqnarray} \label{Ter-wc}
    T^{x,xx}_{a,k}(\omega)&=& (i\hbar)^{-1}V^{x,xx}_{a,k}.
\end{eqnarray}
Furthermore, the leading contribution to the even-order scattering matrix (Fig.~\ref{Fig-tdgr}~(a) and (b)) comes from the self-energy, and according to Eqs.~(\ref{TF-intra}) and (\ref{Selfe-eq}), its single- and bi-exciton components become
\begin{eqnarray}
    \label{TraX-wc}
    T^{x}_{ab}(\omega)&=& i\sum_{k\geq1}\frac{V^{x,xx}_{a,k}V^{xx,x}_{k,b}}
            {\hbar^2(\omega-\tilde\omega^{xx}_{k})},
    \\\label{TraXX-wc}
    T^{xx}_{kl}(\omega)&=& i\sum_{a\geq0}\frac{V^{xx,x}_{k,a}V^{x,xx}_{a,l}}
            {\hbar^2(\omega-\tilde\omega^{x}_{a})},
\end{eqnarray}
respectively.

To calculate the time-dependent Green function, we substitute Eqs.~(\ref{g-wc})--(\ref{TraXX-wc}) into Eqs.~(\ref{GT-rep})--(\ref{T-mtrx}). Further use of Fourier transformation (Eq.~(\ref{Gf-Four})) leads to the following expressions:
\begin{eqnarray}
\label{GtX-wc}
    G^{x}_{ab}(t)&=& \delta_{ab}e^{-i\bar\omega^x_{a}t}
    +\Lambda^{x}_{ab}\left(e^{-i\bar\omega^{x}_at}-e^{-i\bar\omega^x_bt}\right)
    \\\nonumber &-&
    \sum_{k\geq1}\Lambda^{x,xx}_{a,k}\Lambda^{xx,x}_{k,b}e^{-i\bar\omega^{xx}_kt},
\\\label{GtXX-wc}
    G^{xx}_{kl}(t)&=& \delta_{kl}e^{-i\bar\omega^{xx}_{k}t}
    +\Lambda^{xx}_{kl}\left(e^{-i\bar\omega^{xx}_kt}-e^{-i\bar\omega^{xx}_lt}\right)
    \\\nonumber &-&
    \sum_{a\geq0}\Lambda^{xx,x}_{k,a}\Lambda^{x,xx}_{a,l}e^{-i\bar\omega^{x}_at},
\\\label{GtCM-wc}
    G^{x,xx}_{a,k}(t)&=&\Lambda^{x,xx}_{a,k}\left(e^{-i\bar\omega^x_at}-e^{-i\bar\omega^{xx}_kt}\right).
\end{eqnarray}
Here, the shorthand notations for the renormalized complex single- and bi-exciton {\it quasiparticle} frequencies 
\begin{eqnarray}
\label{wx-qprt}
\bar\omega^x_a &=& \tilde\omega^x_a+\sigma^x_a
\\\label{wxx-qprt} 
\bar\omega^{xx}_k &=& \tilde\omega^{xx}_k+\sigma^{xx}_k,  
\end{eqnarray}
are used,  respectively. They contain the following self-energy corrections
\begin{eqnarray}
    \label{SE-X}
    \sigma^{x}_{a}&=&\sum_{k\geq1}\frac{V^{x,xx}_{a,k}V^{xx,x}_{k,a}}
            {\hbar^2\left(\tilde\omega^{x}_a-\tilde\omega^{xx}_k\right)}
    \\\label{SE-XX}
    \sigma^{xx}_{k}&=&\sum_{a\geq0}\frac{V^{xx,x}_{k,a}V^{x,xx}_{a,k}}
            {\hbar^2\left(\tilde\omega^{xx}_k-\tilde\omega^{x}_a\right)}.
\end{eqnarray}
Finally, the transition amplitudes in Eqs.~(\ref{GtX-wc})--(\ref{GtCM-wc})  are
\begin{eqnarray}
    \label{Lmbd-X}
    \Lambda^{x}_{ab}&=&(1-\delta_{ab})\sum_{k\geq1}\frac{V^{x,xx}_{a,k}V^{xx,x}_{k,b}}
            {\hbar^2\left(\bar\omega^{x}_a-\bar\omega^{xx}_k\right)
				\left(\bar\omega^{x}_a-\bar\omega^{x}_b\right)}
    \\\label{Lmbd-XX}
    \Lambda^{xx}_{kl}&=&(1-\delta_{kl})\sum_{a\geq0}\frac{V^{xx,x}_{k,a}V^{x,xx}_{a,l}}
            {\hbar^2\left(\bar\omega^{xx}_k-\bar\omega^{x}_a\right)
				\left(\bar\omega^{xx}_k-\bar\omega^{xx}_l\right)}
    \\\label{Lmbd-CM}
    \Lambda^{x,xx}_{a,k}&=& \frac{V^{x,xx}_{a,k}}
            {\hbar\left(\bar\omega^x_a-\bar\omega^{xx}_k\right)}.
\end{eqnarray}

This representation for the time-domain Green function (Eqs.~(\ref{GtX-wc})--(\ref{Lmbd-CM})) is accurate up to second-order terms in the interband Coulomb interactions. In the following, the above expressions are employed to provide the leading contributions to the single- and bi-exciton photogenerated populations and to derive a set of rate equations for the population relaxation.  

\subsection{Single- and bi-exciton photogeneration}
\label{ppwc}

\begin{figure*}[t]
	\centering
	{\includegraphics[width=6.0in]{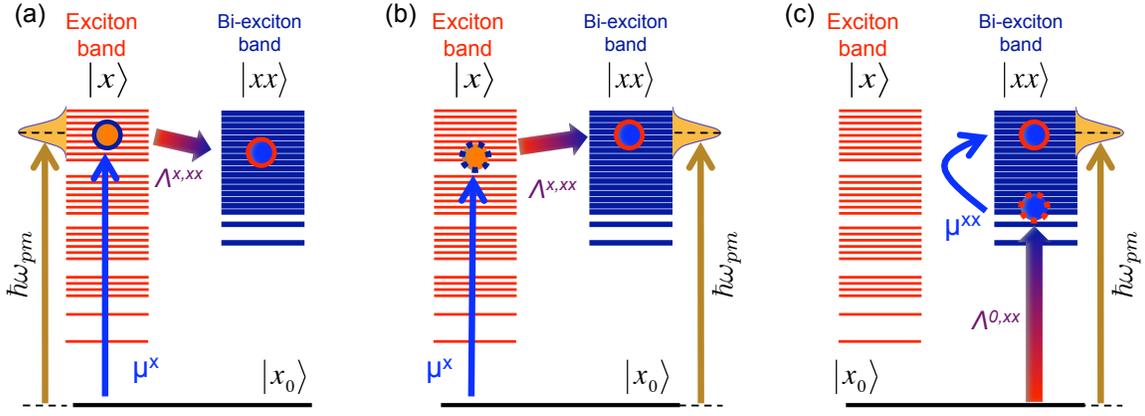}}
	\caption{Bi-exciton photogeneration pathways in the weak Coulomb limit. Panels (a) and (b) show the two components of the pathway involving the vacuum to single-exciton dipole transition ($\mu^{x}$) and the interband Born scattering ($\Lambda^{x,xx}$). In panel~(a), the intraband transition is in resonance with the optical pulse ($\hbar\omega_{pm}$) but the final bi-exciton energy is distributed around $\hbar\omega_{pm}$ according to the non-zero components of $\Lambda^{x,xx}$. Panel~(b) describes the opposite situation, where the single-exciton is virtual and final bi-exciton state is in resonance with the optical pulse. (c) The pathway containing the production of virtual bi-exciton states due to the Born scattering from the exciton vacuum ($\Lambda^{0,xx}$) followed by the intraband dipole transition $\mu^{xx}$. Here, the final bi-exciton state is in resonance with the optical pulse.}
	\label{Fig-spath}
\end{figure*}	

The use of the Green functions represented by Eqs.~(\ref{GtX-wc})--(\ref{Lmbd-CM}) together with Eqs.~(\ref{rho-cp})--(\ref{Ip-eigs}) results in the following form of the photo-generated single-exciton population: 
\begin{eqnarray}
\label{nx-2ord}
    \rho^x_{a} = {n^{x}_a}^{(0)}+{n^{x}_a}^{(1)}+{n^{x}_a}^{(2)}+{{c}^{x}_a}^{(2)},
\end{eqnarray}
where zeroth-, first-, and second-order terms describing the contributions due to the optically prepared quasiparticle populations are: 
\begin{eqnarray}
\label{nx-coef-0}
    {n^{x}_a}^{(0)}&=&|\mu^x_{a0}|^2 {\cal I}(\bar\omega^x_a-\omega_{pm}),
\\\label{nx-coef-1}
    {n^{x}_a}^{(1)}&=&2Re\sum_{k\geq1}{\mu^{x,xx}_{a,k}}
     {\Lambda^{xx,x}_{k,0}}{\mu^x_{0a}}{\cal I}(\bar\omega^x_a-\omega_{pm}),
 \\\label{nx-coef-2}
    {n^{x}_a}^{(2)}&=&
    \left\{\left|\sum_{k\geq1}\mu^{x,xx}_{a,k}{\bar\Lambda^{xx,x}_{k,0}}\right|^2
    +\sum_{k\geq1}\left|\mu^{x}_{a0}\bar\Lambda^{x,xx}_{0,k}\right|^2
   \right.\\\nonumber &+&\left.
   2Re\sum_{b\geq0}\left(\mu^x_{0a}{\Lambda^x_{ab}}\mu^x_{b0}
   +\mu^x_{0a}\mu^x_{ab}{\bar\Lambda^x_{b0}}\right)
 \right.\\\nonumber &+&\left.
   2Re\sum_{kl\geq1}\mu^x_{0a}\Lambda^{x,xx}_{a,k}\mu^{xx}_{kl}\bar\Lambda^{xx,x}_{l,0}
   \right\}
   {\cal I}(\bar\omega^x_a-\omega_{pm}),\;\;\;\;\;
\end{eqnarray}
respectively. They contain the pulse self-convolution function (Eq.~(\ref{Ip-eigs})) which is resonant only at {\em single}-exciton quasiparticle frequencies.   

In contrast, the quasiparticle coherences contributing to $\rho^x_a$ contain both single- and bi-exciton resonances
\begin{eqnarray}\label{nx-coef-3ch}
     {{c}^{x}_a}^{(2)}&=&
   -2Re\sum_{b\geq1}{\mu^x_{0a}}\Lambda^{x}_{ab}{\mu^{x}_{b0}}
   \\\nonumber &\times&
   {\cal I}(\bar\omega^x_b-\omega_{pm};\bar\omega^x_a-\omega_{pm})
   \\\nonumber &-&
    2Re\sum_{k\geq1}\mu^{x}_{0a}\Lambda^{x,xx}_{a,k}
\\\nonumber&\times&
    \left(\sum_{l\geq1}\mu^{xx}_{kl}\bar\Lambda^{xx,x}_{l,0}
+    \sum_{b\geq1}\Lambda^{xx,x}_{k,b}\mu^x_{b0}
    \right)
   \\\nonumber &\times&
   {\cal I}(\bar\omega^{xx}_k-\omega_{pm};\bar\omega^{x}_a-\omega_{pm}),
\end{eqnarray}
entering the pulse self-convolution function (Eq.~(\ref{Ic-eigs})). Although Eqs.~(\ref{nx-coef-0})--(\ref{nx-coef-3ch}) are important for numerical calculations of the QE, we do not discuss the scattering pathways associated with each term, since these pathways carry no information about the CM dynamics. 

The photogenerated bi-exciton population in the weak Coulomb limit according to Eqs.~(\ref{rho-cp})--(\ref{Ip-eigs}) and Eqs.~(\ref{GtX-wc})--(\ref{Lmbd-CM}) is
\begin{eqnarray}
\label{rxx-2ord}
    \rho^{xx}_{k} = {n^{xx}_k}^{(2)}+{{c}^{xx}_k}^{(2)},
\end{eqnarray}
where the quasiparticle population contribution is
\begin{eqnarray}\label{nxx-pl}
    {n^{xx}_k}^{(2)}&=& \sum_{a\geq1}
                \left|\Lambda^{xx,x}_{k,a}\mu^x_{a0}\right|^2{\cal I}(\bar\omega^x_a-\omega_{pm})
 \\\nonumber &+&
    \left|\sum_{a\geq1}\Lambda^{xx,x}_{k,a}\mu^x_{a0}
    +\sum_{l\geq1}\mu^{xx}_{kl}\bar\Lambda^{xx,x}_{l,0}
    \right|^2
    \\\nonumber&\times& 
    {\cal I}(\bar\omega^{xx}_k-\omega_{pm}), 
\end{eqnarray}
and the quasiparticle coherence contribution is
\begin{eqnarray}
\label{nxx-ch}
     {{c}^{xx}_k}^{(2)}&=&
   - 2Re\sum_{(a\neq b)\geq1}{\mu^x_{0a}}^*\Lambda^{x,xx}_{a,k}{\Lambda^{xx,x}_{k,b}}^*{\mu^{x}_{b0}}^*
   \\\nonumber &\times&
   {\cal I}(\bar\omega^x_a-\omega_{pm};\bar\omega^{x}_b-\omega_{pm})
   \\\nonumber &-&
    2Re\sum_{a\geq1}{\mu^{x}_{0a}}^*\Lambda^{x,xx}_{a,k}
    \\\nonumber&\times&
    \left[\sum_{b\geq1}{\Lambda^{xx,x}_{k,b}}^*{\mu^x_{b0}}^*
    +\sum_{l\geq1}{\mu^{xx}_{kl}}^*\bar\Lambda^{xx,x}_{l,0}\right]
    \\\nonumber &\times&
   {\cal I}(\bar\omega^{x}_a-\omega_{pm};\bar\omega^{xx}_k-\omega_{pm}).
\end{eqnarray}
Note that Eqs.~(\ref{rxx-2ord})--(\ref{nxx-ch}) contain only second-order Coulomb terms.

Equation~(\ref{nxx-pl}) has a clear physical interpretation, illustrated in Fig.~\ref{Fig-spath}, where two interfering photogeneration pathways can be distinguished: The first pathway is shown in panels (a) and (b) and both the vacuum to single-exciton dipole transition ($\mu^{x}_{0a}$) and interband Born scattering ($\Lambda^{x,xx}_{a,k}$). The product, $\mu^{x}_{0a}\Lambda^{x,xx}_{a,k}$, of the latter quantities enters the first and second summations over single-exciton index $a$. These sums describe the redistribution of the bare single-exciton oscillator strength between the quasiparticle single-exciton (panel~(a)) and the bi-exciton (panel~(b)) resonances. The second pathway, shown in panel~(c), is represented by the last sum over the bi-exciton index, $l$. Here, the optical transition to the quasiparticle bi-exciton resonance is a combination of the Born scattering event between the vacuum and a bi-exciton state ($\bar\Lambda^{x,xx}_{0,k}$) and  the bi-exciton intraband transition ($\mu^{xx}_{kl}$). Comparison of Eqs.~(\ref{nxx-pl}) and (\ref{nxx-ch}) shows that the latter contains the interference of similar scattering pathways.

\subsection{Population relaxation}
\label{prwc}

To derive a set of rate equations for the population relaxation, we first represent the Redfield equation (Eq.~(\ref{Redfeq})) in the bare single- and bi-exciton basis. In this representation, populations and coherences are coupled. We eliminate the coherences and obtain a memory kernel that depends on the interband Coulomb coupling. We further apply the Markov approximation to the kernel based on the main assumption that the Coulomb interaction is much smaller than the line width arising from the pure dephasing processes. This procedure results in the following set of rate equations:
\begin{eqnarray}
\label{Rate-X}
    \dot \rho^{x}_a &=& -\sum_{m} k^{x,xx}_{a,m}\left(\rho^{x}_a-\rho^{xx}_m\right)
     \\\nonumber&-&
              \sum_{b}\left(\Gamma^{x}_{ba}\rho^{x}_a-\Gamma^{x}_{ab}\rho^{x}_b\right),
    \\\label{Rate-XX}
    \dot \rho^{xx}_k &=&-\sum_{b} k^{x,xx}_{b,k}\left(\rho^{xx}_k-\rho^{x}_b\right)
     \\\nonumber&-&
        \sum_{m}\left(\Gamma^{xx}_{mk}\rho^{xx}_k-\Gamma^{xx}_{km}\rho^{xx}_m\right).
\end{eqnarray}

\begin{figure}[t]
	\centering
	{\includegraphics[width=2.0in]{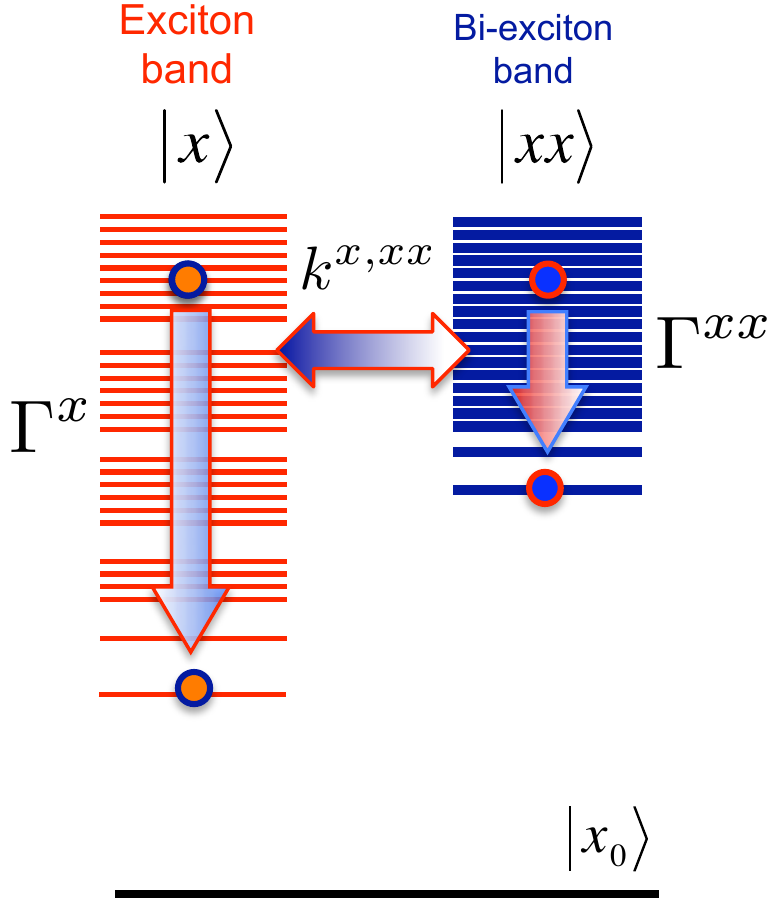}}
	\caption{Population relaxation in the weak Coulomb limit consists of uncoupled interband Auger recombination and 
	impact ionization processes with rate $k^{x,xx}$, and the intraband phonon-induced cooling with rates $\Gamma^x$ 
	and $\Gamma^{xx}$.}
	\label{Fig-wrlx}
\end{figure}

Here, the interband scattering and the intraband phonon-induced population cooling are described by different terms indicating that in the weak Coulomb limit these two processes are uncoupled (Fig.~\ref{Fig-wrlx}). Specifically, the first term in the r.h.s. of each equation describes the interband population transfer due to both the impact ionization and the Auger recombination processes. The related population transfer rate, arising from the Markov kernel appearing in the coherence elimination, is 
\begin{eqnarray}
\label{k-x-xx}
    k^{x,xx}_{a,n} &=& \frac{2}{\hbar^2}\left|V^{x,xx}_{a,k}\right|^2\frac{\gamma^{x,xx}_{a,k}}
            {\left(\omega^{x}_{a}-\omega^{xx}_{k}\right)^2+(\gamma^{x,xx}_{a,k})^2},
\end{eqnarray}
where $\gamma^{x,xx}_{a,n}$ is the pure dephasing rate. 
 
The second terms in the r.h.s. of both Eq.~(\ref{Rate-X}) and (\ref{Rate-XX}), describes phonon-assisted cooling. The entering population decay rates can be obtained from the general expression given by Eq.~(\ref{prlx-rate}) and (\ref{Yqp-o}),  where we set $\bar\Lambda_{ab}(\omega_x) = \delta_{ab}$, $\bar\Lambda_{kl}(\omega_{xx}) = \delta_{kl}$, and $\bar\Lambda_{ak}(\omega_{x,xx}) = 0$. These transition amplitudes arise from the {\em zero}-order Coulomb terms of the Green functions given by Eqs.~(\ref{GtX-wc})--(\ref{GtCM-wc}). The remaining second-order transition amplitudes are dropped, since together with the exciton-phonon couplings their net contributions to the rates become negligibly small. 

Since, during the transformation from the quasiparticle representation back to the bare single- and bi-exciton states, one has to keep only zeroth-order Coulomb terms, the form of the population relaxation rates does not change. As a result, their expressions are
\begin{eqnarray}\label{prlx-rate-X}
    \Gamma^{x}_{ab} &=& \frac{1}{\hbar^2}\sum_{\alpha\alpha^{'}}
        Y^{x}_{ab;\alpha}Y^{x}_{ab;\alpha^{'}} C_{\alpha\alpha^{'}}(\omega^{x}_{a}-\omega^{x}_{b}),
    \\\label{prlx-rate-XX}
    \Gamma^{xx}_{kl} &=& \frac{1}{\hbar^2}\sum_{\alpha\alpha^{'}}
        Y^{xx}_{kl;\alpha}Y^{xx}_{kl;\alpha^{'}} C_{\alpha\alpha^{'}}(\omega^{xx}_{k}-\omega^{xx}_{l}),
\end{eqnarray}
where the phonon correlation function, $\tilde C_{\alpha\alpha^{'}}(\omega)$, is defined in Sec.~\ref{psrm}, and the exciton-phonon coupling constants, $Y^{x}_{ab;\alpha}$ and $Y^{xx}_{kl;\alpha}$, are defined by Eqs.~(\ref{Y-X}) and (\ref{Y-XX}), respectively.\cite{ftn-07} Similarly, the following expression for the pure dephasing rate immediately follows from Eqs.~(\ref{dphas-rate}), (\ref{Yqp-d}), (\ref{GtX-wc}), and (\ref{GtXX-wc}) 
\begin{eqnarray}\label{dphas-rate-X-XX}
    \gamma^{x,xx}_{a,n} &=&\frac{1}{\hbar^2}\sum_{\alpha\alpha^{'}}
    \left(Y^{x}_{aa;\alpha}-Y^{xx}_{nn;\alpha}\right)
  \\\nonumber&\times&
    \left(Y^{x}_{aa;\alpha^{'}}-Y^{xx}_{nn;\alpha^{'}}\right)
        \tilde C^{'}_{\alpha\alpha^{'}}(0).
\end{eqnarray}

Finally, we outline the computation of the QE in the weak Coulomb limit. First, Eqs.~(\ref{nx-2ord})--(\ref{nxx-ch}) are evaluated to find the initial conditions, $\rho^{x}_a(0)$ and $\rho^{xx}_k(0)$, for the population relaxation. These density matrix elements can also be used to obtain the QE (Eq.~(\ref{QE})) associated with the photogeneration processes. Next, starting with the latter boundary conditions, the wavepackets, $\rho^{x}_a(t)$ and $\rho^{xx}_k(t)$, should be numerically propagated to the bottom of the single- and bi-exciton bands according to Eqs.~(\ref{Rate-X}) and (\ref{Rate-XX}) with the parameters defined by Eqs.~(\ref{k-x-xx})--(\ref{dphas-rate-X-XX}). This provides input for the determination of the total QE.

\section{Discussion}
\label{disc}

In this section, we discuss the relation between our proposed Exciton Scattering Model and three earlier models: the Coherent Superposition Model,\cite{shabaev06} the Direct Photogeneration Model,\cite{schaller05,rupasov07} and the Impact Ionization Model\cite{franceschetti06,allan06}. 

\subsection{Coherent Superposition Model}
\label{csm}

\begin{figure}[t]
	\centering
	{\includegraphics[width=3.4in]{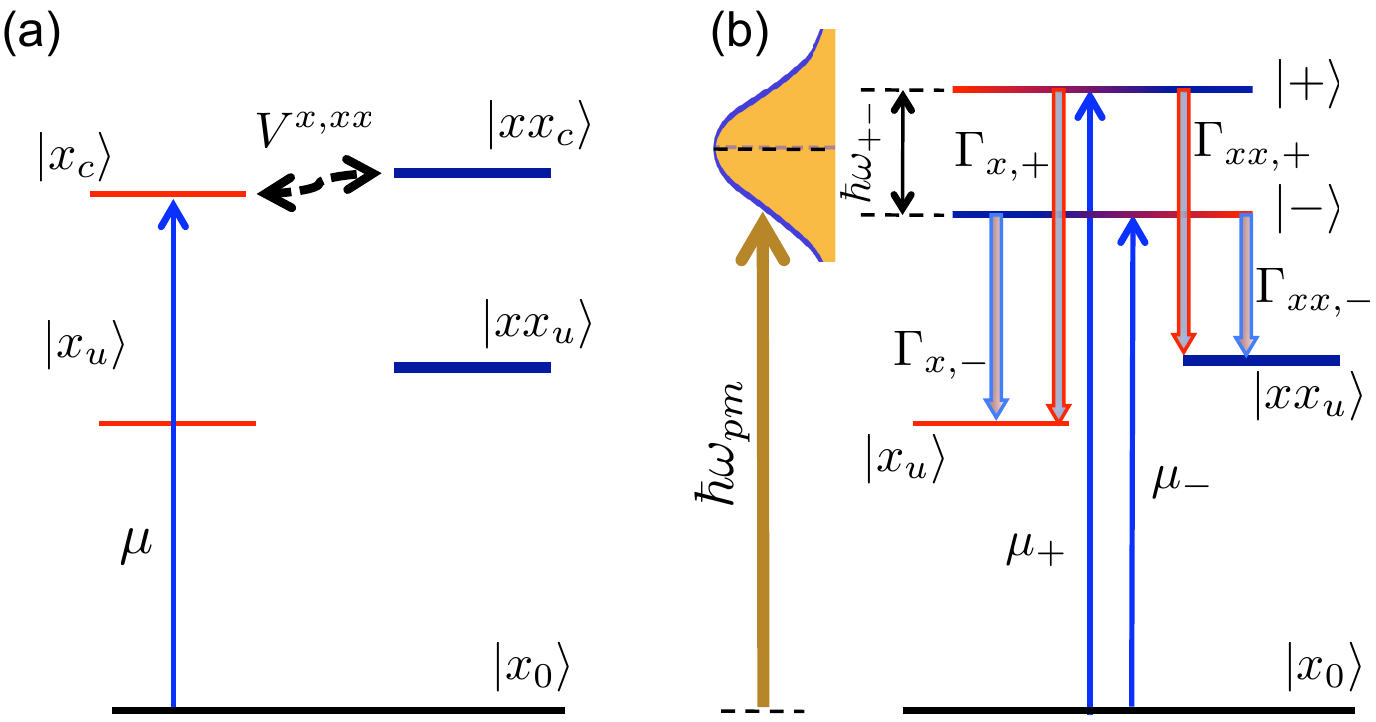}}
	\caption{Level diagram for the Coherent Superposition model: (a) Bare single- and bi-exciton state representation where
	two states $|x_{c}\rangle$ and $|xx_{c}\rangle$ are coupled by Coulomb matrix element $V^{x,xx}$, and two states 
	$|x_{u}\rangle$ and $|xx_{u}\rangle$ are uncoupled. The transition dipole, $\mu$, couples the vacuum state with the upper single-exciton states.
		(b) Quasiparticle representation. Scattering processes correspond to the optical transitions with 
		$\mu_+=\sqrt{\Lambda^x_+}\mu$ and $\mu_-=\sqrt{\Lambda^x_-}\mu$. In a short-pulse limit, the pump spectral width
		exceeds $\hbar\omega_{+-}$, and both the coherence and populations of $|\pm\rangle$ states are
		included. Their projections back to the bare single- and bi-exciton populations are given by Eqs.~(\ref{rxc-sp}) 
		and (\ref{rxxc-sp}). The population relaxation pathways with rates entering Eqs.~(\ref{rxc-sp})--(\ref{rxxu-sp})
		are shown by the arrows pointed down.} 
	\label{Fig-csm}
\end{figure}

The Coherent Superposition Model is the limit of our Exciton Scattering Model, in which only the two states $|x_{c}\rangle$  and $|xx_{c}\rangle$ are coupled by the Coulomb matrix element $V^{x,xx}$, and the two states $|x_{u}\rangle$ and $|xx_{u}\rangle$ are decoupled as shown in Fig.~\ref{Fig-csm}~(a). There is also no coupling to the vacuum state. The $|x_{c}\rangle$  and $|xx_{c}\rangle$ states are assumed to be almost degenerate, i.e. $\hbar\omega^{xx}_c\sim\hbar\omega^x_c$, leading to the strong interaction condition $\hbar|\omega^{xx}_c-\omega^x_c|\ll V^{xx,x}$. The details of the calculations of the scattering matrix components, the single- and bi-exciton Green functions in the framework of the Coherent Superposition Model are given in Appendix~\ref{cspm}. According to these calculations, strong Coulomb interaction corresponds to the splitting between coupled states, and formation of the quasiparticle states $|\pm\rangle$ with energies, $\hbar\omega_\pm$ (Eq.~(\ref{wpm})), as illustrated in Fig.~\ref{Fig-csm}~(b). Another assumption used in the calculations is that the splitting, $\omega_{+-}=\omega_+-\omega_-$, significantly exceeds the quasiparticle level broadening. These assumptions in our new notations reproduce the model proposed in Ref.~\onlinecite{shabaev06}

The authors of Ref.~\onlinecite{shabaev06} used the phonon-assisted relaxation model containing uncoupled intraband relaxation pathways for single- and bi-excitons resulting in independent cooling within each manifold. We argue that these relaxation pathways are {\em coupled} since the Coulomb interaction is strong.\cite{kubo95} As a result the interband phonon assisted processes should be accounted for. Furthermore, we point out in Sec.~\ref{psrm}, that the relaxation equations should reproduce the equilibrium distribution function not for the bare single- and bi-exciton states but for the quasiparticle states.  Another assumption used in Ref.~\onlinecite{shabaev06} is that the dephasing rate between coupled states is fully determined by the population relaxation processes. However, the {\em pure dephasing} time in NCs is estimated to be several orders of magnitude shorter than the population relaxation time.\cite{kamisaka06,kamisaka08} In this case, the dephasing rate is totally due to the pure dephasing, indicating separation of the timescales for the coherence and population dynamics. 

The population relaxation rates from the quasiparticle states to the uncoupled single- and bi-exciton states, shown in Fig.~\ref{Fig-csm}~(b), immediately follow from Eqs.~(\ref{prlx-rate}) and (\ref{Yqp-o}) with the transition amplitudes $\Lambda^{x}_{\pm}$ and $\Lambda^{xx}_{\pm}$ (Eqs.~(\ref{LmX}) and (\ref{LmXX})) inserted. The rates are
\begin{eqnarray}\label{prx-cw}
    \Gamma_{x,\pm} &=& \frac{1}{\hbar^2}\sum_{\alpha}[Y^x_{uc;\alpha}]^2\Lambda^x_{\pm}
    				C_{\alpha}(\omega^x_u-\omega_{\pm}),
\\\label{prxx-cw}
 \Gamma_{xx,\pm} &=& \frac{1}{\hbar^2}\sum_{\alpha}
        [Y^{xx}_{uc;\alpha}]^2\Lambda^{xx}_{\pm}C_{\alpha}(\omega^{xx}_u-\omega_{\pm}),
\end{eqnarray}
where $Y^x_{uc;\alpha}$ and $Y^{xx}_{uc;\alpha}$ are the exciton-phonon interactions connecting the uncoupled and coupled states. Note  that these expressions reproduce the uncoupled single- and bi-exciton relaxation rates only if the Coulomb interaction is weak, $\hbar|\omega^{xx}_c-\omega^x_c|\gg V^{xx,x}$.\cite{ftn-08} Population transfer also exists {\em between} the quasiparticle states. However, for the sake of simplicity, we drop this pathway. The derivation of the pure dephasing rate, $\gamma_{+-}$, between the quasiparticle states is discussed in Appendix~\ref{cspm}.
 
The Coherent Superposition Model consists of only one optical transition induced by the dipole moment, $\mu$, between the vacuum and coupled single-exciton state shown in panel (a) of Fig.~\ref{Fig-csm}. The multiple-scattering processes redistribute the oscillator strength so that both quasiparticle states become optically allowed as shown in panel~(b). To observe the oscillations of the pump-probe signal (bleach) predicted by the Coherent Superposition Model, the pulse duration should be less than the dephasing time and spectral widths of the pulse should exceed the level splitting, $\hbar\omega_{+-}$. This condition is satisfied in the so-called impulsive limit in which  the pulse self-convolution function becomes frequency-independent\cite{ftn-09} 
\begin{eqnarray}
\label{Isp}
    {\cal I}_{sp}\equiv {\cal I}_{sp}(\tilde\omega_{\pm}-\omega_{pm})&=& 
 \\\nonumber{\cal I}_{sp}(\tilde\omega_{+}-\omega_{pm};\tilde\omega_{-}-\omega_{pm})&=&
4\hbar^{-2}\bar\tau_{pm}^2[{\cal E}^{(0)}_{pm}]^2.
\end{eqnarray}
Here, ${\cal E}^{0}_{pm}$ is the amplitude of the pump pulse, and $\bar\tau_{pm}=\sqrt{\pi}\tau_{pm}$ is the effective pulse duration.     

To find the time-dependent populations of the coupled states prepared by a short pulse, we use the expressions for the corresponding density matrix components (Eqs.~(\ref{nxc}), (\ref{cxxc}), (\ref{nxxc}), and (\ref{cxxc})) obtained in Appendix~\ref{cspm} along with Eq.~(\ref{Isp}). This results in the following populations of coupled single- and bi-exciton states:
\begin{eqnarray}
\label{rxc-sp}
    \rho^x_c(\tau) &=& A^2 \sum_{\bar\xi=\pm}[\Lambda^x_{\bar\xi}]^2 
    e^{-\left(\Gamma_{x,\bar\xi}+\Gamma_{xx,\bar\xi}\right)\tau}
\\\nonumber &+&
    2A^2\Lambda^{x}_{+}\Lambda^{x}_{-}\cos(\omega_{+-}\tau)e^{-\gamma_{+-}\tau},
\\\label{rxxc-sp}
    \rho^{xx}_c(\tau) &=& A^2 \sum_{\bar\xi=\pm} [\Lambda^{x,xx}_{\bar\xi}]^2 
    e^{-\left(\Gamma_{x,\bar\xi}+\Gamma_{xx,\bar\xi}\right)\tau}
\\\nonumber \\\nonumber &-&
    2A^2\Lambda^{x}_{+}\Lambda^{x}_{-}\cos(\omega_{+-}\tau)e^{-\gamma_{+-}\tau},
\end{eqnarray}
respectively. The populations of the uncoupled states as a function of delay time, $\tau$, can be obtained from Eqs.~(\ref{nxu}) and (\ref{nxxu}) together with Eq.~(\ref{Isp}):  
\begin{eqnarray}
\label{rxu-sp}
    \rho^x_u(\tau) &=& A^2 \sum_{\bar\xi=\pm} \frac{\Gamma_{x,\bar\xi}\Lambda^x_{\bar\xi}}
    {\Gamma_{x,\bar\xi}+\Gamma_{xx,\bar\xi}}
\\\nonumber&\times &    
    \left\{1-e^{-\left(\Gamma_{x,\bar\xi}+\Gamma_{xx,\bar\xi}\right)\tau}\right\}, 
\\\label{rxxu-sp}
    \rho^{xx}_u(\tau) &=& A^2 \sum_{\bar\xi=\pm} \frac{\Gamma_{xx,\bar\xi}\Lambda^x_{\bar\xi}}
    {\Gamma_{x,\bar\xi}+\Gamma_{xx,\bar\xi}}
\\\nonumber&\times &    
	    \left\{1-e^{-\left(\Gamma_{x,\bar\xi}+\Gamma_{xx,\bar\xi}\right)\tau}\right\}.
\end{eqnarray}
Here, $A = 2\mu{\cal E}^{(0)}\bar\tau_{pm}/\hbar$ is a dimensionless parameter; $\Lambda^{x}_{\bar\xi}$,  $\Lambda^{x,xx}_{\bar\xi}$, and $\Lambda^{xx}_{\bar\xi}$ are the transition amplitudes defined in Eqs.~(\ref{LmX})--(\ref{LmInt}).

The oscillations predicted by the Coherent Superposition Model exist only between coupled states, and have a frequency equal to the quasiparticle level splitting, $\omega_{+-}$. Specifically, they are described by the second term in Eqs.~(\ref{rxc-sp}) and (\ref{rxxc-sp}). Since there is no oscillating term in the populations of the uncoupled states, oscillations of the bleach can be observed in experiments in which the probe directly monitors the time evolution of $\rho^{x}_c$ and/or $\rho^{xx}_c$. In fact, the expression for bleach, given by Eq.~(7) in Ref.~\onlinecite{shabaev06}, contains non-vanishing contributions from the coupled bi-exciton population.

Finally, we compare the QE of the photogeneration event, and the total QE after population cooling. The photogeneration QE can be determined from Eqs.~(\ref{QE}), (\ref{rxc-sp}), and (\ref{rxxc-sp}) on timescales longer than the dephasing occurs but shorter than the population relaxation. This quantity is:  
\begin{eqnarray}
\label{QE0}
    QE_<=\frac{12(V^{x,xx})^2+(\omega^{xx}_c-\omega^x_c)^2}{4(V^{x,xx})^2+(\omega^{xx}_c-\omega^x_c)^2}.
\end{eqnarray}
It contains no relaxation parameters. As the strength of the Coulomb coupling increases, $QE_<$ approaches its maximum value of $3/2$.  

The total QE can be determined from Eqs.~(\ref{QE}), (\ref{rxu-sp}), and (\ref{rxxu-sp}) in the limit in which time is longer than the typical population relaxation time. This results in the sum,
\begin{eqnarray}
\label{QEi}
    QE_>=\sum_{\bar\xi=\pm}\Lambda^x_{\bar\xi}\frac{2\Gamma_{xx,\bar\xi}+\Gamma_{x,\bar\xi}}
    			{\Gamma_{xx,\bar\xi}+\Gamma_{x,\bar\xi}},
\end{eqnarray}
where $\Lambda^x_{\bar\xi}$ determines the probability of optical excitation for each quasiparticle state, $\bar\xi=\pm$, and the ratio of the population relaxation rates gives the maximum QE associated with each of the states. (Fig.~\ref{Fig-csm}~(b)). We emphasize, that this expression for QE accounts for the relaxation pathways mixing the coupled and uncoupled states of different multiplicities. The latter contributions have not been considered before and are expected to be non-negligible. As the Coulomb interaction increases, $QE_>$ approaches its maximum value given by the ratio of the {\em total} population relaxation rates $(2\Gamma^{xx}+\Gamma^{x})/(\Gamma^{xx}+\Gamma^{x})$, where $\Gamma^{x} = \Gamma_{x,+} = \Gamma_{x,-}$ and  $\Gamma^{xx} = \Gamma_{xx,+} = \Gamma_{xx,-}$.\cite{ftn-10} Regardless the differences in the relaxation models, similar maximum values for $QE_>$ were obtained in Ref.~\onlinecite{shabaev06} for increasing Coulomb interactions.

\subsection{Direct Photogeneration and Impact Ionization Models}
\label{dpii}

The Direct Photogeneration Model assumes weak Coulomb coupling between single- and bi-exciton states. Therefore, to find the bi-exciton generation rate, we should begin with the expressions for the photoinduced bi-exciton population (Eq.~(\ref{rxx-2ord})--(\ref{nxx-ch})) derived in Sec.~\ref{ppwc}. An additional assumption of the model is that the pump pulse is much longer than the dephasing times, resulting in the so-called continuous wave (CW) limit. The pulse self-convolution function for quasiparticle populations in this CW limit becomes proportional to  the Lorentzian line-shape function:\cite{ftn-11}
\begin{eqnarray}
\label{ICW}
    {\cal I}_{CW}(\tilde\omega_{\bar\xi g}-\omega_{pm})&=&\frac{2}{\hbar^2}
    \frac{{{\cal E}^{(0)}_{pm}}^2\bar\tau_{pm}\gamma_{\bar\xi g}} {(\omega_{\bar\xi g}-\omega_{pm})^2+(\gamma_{\bar\xi g})^2},
\end{eqnarray}
where ${\cal E}^{0}_{pm}$ is the pump pulse electric field amplitude, and $\bar\tau_{pm}=\sqrt{\pi}\tau_{pm}$ is the effective pulse duration. The pulse self-convolution function associated with the quasiparticle coherences vanishes in CW limit. 

Next, we introduce the bi-exciton generation rate as $W_{xx}=\sum_{k\geq 1}\rho^{xx}_{k}/\bar\tau_{pm}$ where the limit of $\gamma_{\bar\xi 0}\rightarrow 0$ should be taken. This, according to Eqs.~(\ref{rxx-2ord}) and (\ref{nxx-ch}), corresponds to the following expression for the bi-exciton generation rate:
\begin{eqnarray}\label{Wxx-2ord}\nonumber
    W_{xx}&=&\frac{2\pi}{\hbar}{{\cal E}^{(0)}_{pm}}^2 \sum_{k\geq1}\sum_{a\geq1}
                \left|\Lambda^{xx,x}_{k,a}\mu^x_{a0}\right|^2\delta(E^x_a-\hbar\omega_{pm})
 \\\nonumber &+&
    \frac{2\pi}{\hbar} {{\cal E}^{(0)}_{pm}}^2\sum_{k\geq1}\left|\sum_{a\geq1}\Lambda^{xx,x}_{k,a}\mu^x_{a0}
    +\sum_{l\geq1}\mu^{xx}_{kl}\bar\Lambda^{xx,x}_{l,0}
    \right|^2
    \\&\times& 
    \delta(E^{xx}_k-\hbar\omega_{pm}).
\end{eqnarray}
Here, the interband transition amplitude, $\Lambda^{xx,x}_{k,a}$, is given by Eq.~(\ref{Lmbd-CM}), and $E^x_a = \hbar\omega_{a}^{x}$ and $E^{xx}_k = \hbar\omega_{k}^{xx}$ are the single- and bi-exciton energies, respectively.\cite{ftn-12} The leading term in the single-exciton generation rate can be easily obtained using the same approach: 
\begin{eqnarray}\label{Wx-2ord}
    W_{x}&=&\frac{2\pi}{\hbar}{{\cal E}^{(0)}_{pm}}^2 \sum_{a\geq1}
    	\left|\mu^x_{a0}\right|^2\delta(E^x_a-\hbar\omega_{pm}),
\end{eqnarray}
and it coincides  with that given in Ref.~\onlinecite{schaller05}.

Equation~(\ref{Wxx-2ord}) should now be compared with Eqs.~(1)~and~(3) describing the direct bi-exciton generation rate via virtual single-exciton states and via coupling to the vacuum states derived in Refs.~\onlinecite{schaller05} and \onlinecite{rupasov07}, respectively. This comparison shows that the two contributions to the second term in Eq.~(\ref{Wxx-2ord}) multiplied by $\delta\left(E^{x}_{a}-E^{xx}_{k}\right)$ and shown in panels~(b) and (c) of Fig.~\ref{Fig-spath}, reproduce Eqs.~(1)~and~(3) from Refs.~\onlinecite{schaller05} and \onlinecite{rupasov07}, respectively. In summary, the weak Coulomb coupling limit of our Exciton Scattering Model in the particular case of CW excitation not only recovers the previously developed Direct Photogeneration Model but also predicts an additional contribution given by the first term in Eq.~(\ref{Wxx-2ord}) and illustrated in Fig.~\ref{Fig-spath} (a) as well as the interference of the previously studied pathways (Fig.~\ref{Fig-spath}~(b) and (c)) which is clearly seen in the second term of Eq.~(\ref{Wxx-2ord}). 

The central objective of the Impact Ionization Model is the calculation of impact ionization and Auger recombination rates, which can be easily obtained from Eq.~(\ref{k-x-xx}) by taking the limit of $\gamma^{x,xx}_{a,k}\rightarrow 0$, and further performing the summation over the final bi- and single-exciton states
\begin{eqnarray}
\label{WII}
    W_{a}^{II} &=& \frac{2\pi}{\hbar}\sum_{k\geq1}\left|V^{x,xx}_{a,k}\right|^2
    \delta\left(E^{x}_{a}-E^{xx}_{k}\right),
\\\label{WAR}
	W_{k}^{AR} &=& \frac{2\pi}{\hbar}\sum_{a\geq1}\left|V^{x,xx}_{a,k}\right|^2
    \delta\left(E^{x}_{a}-E^{xx}_{k}\right),
\end{eqnarray}
respectively. Comparison of these expressions with Eqs.~(1) and (2) from Ref.~\onlinecite{franceschetti06} leads to the conclusion that the Impact Ionization Model is just the relaxation component of our model in the weak Coulomb coupling regime (Fig.~\ref{Fig-wrlx}). Usually, the initial condition for the impact ionization dynamics is taken to be only the photogenerated {\em single}-exciton population (Eq.~(\ref{nx-coef-0})). We argue that the bi-exciton population described by Eqs.~(\ref{rxx-2ord})--(\ref{nxx-ch}) should also be included, since this contribution is of the same order of magnitude.

The discussion above shows that the Direct Photogeneration Model and the Impact Ionization Model complement each other, and are a particular case of our more general Exciton Scattering Model in the weak Coulomb limit. Specifically, the Direct Photogeneration Model describes the primary photoexcitation process involving a pump pulse that is longer than the dephasing time and shorter than the inverse relaxation rates of impact ionization (Eq.~(\ref{WII})), Auger recombination  (Eq.~(\ref{WAR})) and phonon-assisted decay (Eqs.~(\ref{prlx-rate-X})--(\ref{prlx-rate-XX})). The Impact Ionization Model describes photogenerated population relaxation with the initial conditions given by Eqs.~(\ref{Wxx-2ord})--(\ref{Wx-2ord}). However, we argue that a systematic computational approach should follow from the weak Coulomb limit computational scheme given in Sec.~\ref{wclm}, since this approach contains additional contributions not considered before and also accounts for finite-time pulse excitation.   

\section{Concluding Remarks}
\label{conc}

Currently, numerical implementation of our Exciton Scattering Model is a challenging task, since the calculations involve a large ($\sim 10^5$) number of bi-exciton states. This poses difficulties for the full scattering matrix calculations (Eqs.~(\ref{TF-intra})--(\ref{TF-inter})) associated with large computer memory requirement for the matrix inversion. These difficulties can be  overcome by noticing that in NCs, the energy difference between most of the coupled single- and bi-exciton states is larger than the interband Coulomb interaction, and only a small number of these states are in resonance (degenerate). As a result, the expressions obtained in the weak Coulomb limit (Sec.~\ref{wclm}) should be used to evaluate the contributions from well-separated states, and only contributions from the degenerate states need to be included in the multiple-scattering formalism. If the level broadening for the degenerate states in NCs {\em exceeds} the Coulomb coupling then the full computational scheme developed in Sec.~\ref{wclm} should be used, including the degenerate states.   

The comparison of CM processes in both NCs and in bulk semiconductors is important for understanding the role of quantum size effects on QE. Therefore, we emphasize that our proposed formalism is valid for CM in bulk semiconductors. The transition is simple: One has to replace all the summations over the single- and bi-exciton indices as well as over the quasiparticle states by summations over their quasi-momenta and spin degrees of freedom. All matrix elements entering the calculations can be represented in the quasi-momentum basis set. This representation will automatically impose the quasi-momentum conservation restrictions. 

To summarize, we have proposed the Exciton Scattering Model which treats the two main processes of CM,  photogeneration and population relaxation, on the same footing. Our model is valid in the neighborhood of the AET where the contribution of the higher-multiplicity exciton states (tri-exciton, {\it etc.}) can be neglected. Our model includes relatively large Coulomb interactions leading to multiple interband scattering events. The only restriction on the Coulomb interaction strength is that it should not mix the higher multiplicity states. Based on our general formalism, expressions determining the QE in the limit of weak Coulomb interaction have been derived. This limit is extremely useful for numerical calculations for specific materials. Since the AET is sensitive to the material-dependent selection rules imposed on the Coulomb matrix elements and transition dipole matrix elements, its determination can be done through direct numerical calculations. As we demonstrated, our Exciton Scattering Model recovers three previously proposed models as limiting cases. By including additional mechanisms of CM, our model provides a unified approach to the study of CM in NCs  and in the bulk limit.

\acknowledgements

This work was supported by the Office of Basic Energy Sciences, US Department of Energy, and Los Alamos LDRD funds. We also acknowledge the support provided by CNLS. We wish to thank Victor I. Klimov, Vladimir Chernyak, Sergei Tretiak, Gary D. Doolen, and Darryl L. G. Smith for stimulating discussions and comments on the manuscript. 

\appendix

\section{Interacting multi-exciton Hamiltonian}
\label{EffH}

After introducing the electron and hole creation (annihilation) operators $c^\dag_n$ ($c_n$) and $d^\dag_m$ ($d_m$), respectively, the many-body Hamiltonian describing the valence and conduction band electronic states in semiconductors can be represented as a sum of three components\cite{axt98}
\begin{eqnarray}\label{H_many_body}
    \hat{\cal H}_{eh} = \hat{\cal H}_0 + \hat{\cal H}_1 +\hat{\cal H}_2.
\end{eqnarray}
In this expression, the first term
\begin{eqnarray}\label{Heh}
    \hat {\cal H}_{0} &=& \sum_{m} \epsilon^e_m c_m^\dag c_m
    -\sum_{n} \epsilon^h_n d_n^\dag d_n
\\\nonumber
    &-& \sum_{mnlk} (V^{ehhe}_{mnlk}-V^{eheh}_{mnkl})c_m^\dag d_l^\dag d_n c_k
\\\nonumber
     &+&\frac{1}{2}\sum_{mnlk} V^{eeee}_{mnlk}c_m^\dag c_n^\dag c_l c_k
\\\nonumber
    &+&\frac{1}{2}\sum_{mnlk} V^{hhhh}_{mnlk}d_l^\dag d_k^\dag d_m d_n,
\end{eqnarray}
conserving number of quasiparticles describes those non-interacting electrons and holes which are characterized by the Hartree-Fock energies $\epsilon^e_n$ and $\epsilon^h_n$, respectively. Here and below (Eqs.~(\ref{Heh})--(\ref{H2})), the superscripts $e$ and $h$ denote electrons and holes, respectively, while the corresponding indices run over all Hartree-Fock states in the valence and conduction bands. In  Eq.~(\ref{Heh}) the Coulomb matrix element is
\begin{equation}
 \label{Weeee}
    V_{mnlk}=\int d^3x\int d^3 y\:\psi^*_m(x)\psi^*_n(y)V(|x-y|)\psi_l(y)\psi_k(x),
\end{equation}
where $V(|x-y|)$ is the Coulomb potential and $\psi_n$ is the electron (hole)
Hartree-Fock wave function.

 The second term in the Hamiltonian~(\ref{H_many_body}),
\begin{eqnarray}\label{H1}
    \hat{\cal H}_{1} &=& \frac{1}{2}\sum_{mnlk} (V^{ehee}_{mnlk}-V^{ehee}_{mnkl})
	 c_m^\dag d_n c_l c_k \nonumber \\
         &+&\frac{1}{2}\sum_{mnlk} (V^{eehe}_{mnlk}-V^{eeeh}_{mnkl})
	 c_m^\dag c_n^\dag d_l^\dag c_k \nonumber \\
         &+&\frac{1}{2}\sum_{mnlk}  (V^{hhhe}_{mnlk}-V^{hheh}_{mnkl})
	 d_l^\dag d_m d_n c_k \nonumber \\
         &+&\frac{1}{2}\sum_{mnlk} (V^{ehhh}_{mnlk}-V^{ehhh}_{mnkl})
	 c_m^\dag d_l^\dag d_k^\dag d_n
\end{eqnarray}
describes the processes of creation or annihilation of a single electron-hole pair (in the presence of another electron or hole state) which are referred to as Auger recombination and impact ionization, respectively. Finally, the last term in the Hamiltonian~(\ref{H_many_body})
\begin{eqnarray}\label{H2}
    \hat{\cal H}_{2} &=& \frac{1}{2}\sum_{mnlk} V^{hhee}_{mnlk}d_m d_n c_l c_k
    \nonumber \\
    &+&\frac{1}{2}\sum_{mnlk} V^{eehh}_{mnlk} c_m^\dag c_n^\dag d^\dag_l d^\dag_k
\end{eqnarray}
characterizes those processes which involve the simultaneous creation or annihilation of two electron-hole pairs.

Neglecting all possible charged states, we consider only the space spanned by all possible multiple electron-hole pairs (multi-excitons) $S=\oplus_{\bar n\geq0}S^{\bar n}$ where $S^{0}$ is the exciton vacuum (filled valence band, empty conduction band), and  $S^{\bar n}$ is the exciton space of multiplicity $\bar n$ with the complete basis set constructed from non-interacting electron-hole states
\begin{eqnarray}\label{eh-bas}
|e^{\bar n}_{\bar a}h^{\bar n}_{\bar b}\rangle = \Pi_{k=1}^{\bar n} c^\dag_{a_k} d^\dag_{b_k}|0\rangle,
\end{eqnarray}
where the generalized indices are defined as $\bar a=\{a_1,\dots ,a_{\bar n}\}$ and $\bar b=\{b_1,\dots ,b_{\bar n}\}$. In this representation the many-body Hamiltonian term $\hat{\cal H}_0$ (Eq.~(\ref{Heh})) maps $S^{\bar n}$ on itself. As a result, one can define the eigenstates $|X_{\xi}^{\bar n}\rangle$ of $\hat{\cal H}_0$ forming a complete basis set in $S^{\bar n}$ and, respectively, the eigenenergies, $\hbar\omega^{\bar n}_\xi$. The latter eigenstates describe bound n-exciton states in $S^{\bar n}$ whose eigenenergies include the binding energy due to the electron-electron, hole-hole and electron-hole Coulomb correlations. The introduced bound n-exciton states are related to the non-ineracting electron-hole basis through the unitary transformation
\begin{eqnarray}
\label{multi-X-bas}
 |x_{p}^{\bar n}\rangle = \sum U^{\bar n}_{p;\bar a \bar b}|e_{\bar a}^{\bar n} h_{\bar b}^{\bar n}\rangle,
\end{eqnarray}
defined by the matrix $\{U^{\bar n}_{p;\bar a \bar b}\}$.

The eigenstate equations
\begin{eqnarray}\label{U1-opr}
 \sum_{pq}\langle h_s e_r|\hat {\cal H}_0 - \hbar\omega^x_{a}|e_p h_q\rangle U^x_{a;pq}=0,
\end{eqnarray}
and
\begin{eqnarray}\label{U2-opr}
 \sum_{pqrs}\langle h_e e_f  h_g e_h|\hat {\cal H}_0 - \hbar\omega^{xx}_{n}|e_p h_q e_r h_s\rangle U^{xx}_{n;pqrs}=0,
 \end{eqnarray}
determining the transformation matrix elements for the single- and bi-exciton states, have the following form:
\begin{eqnarray}\label{U1}
   &~&\sum_{pq}\left\{\left(\epsilon^e_p-\epsilon^h_s-
   \hbar\omega^x_a\right)\delta_{sq}\delta_{rp} \right.\nonumber \\
   &+&\left. V^{eheh}_{rqps}-V^{ehhe}_{rqsp}\right\}U^{x}_{a,pq}=0,
\end{eqnarray}
and
\begin{eqnarray}\label{U2}
   &~&\sum_{pqrs}\left\{
   \left(\delta_{gq}\delta_{es}-\delta_{eq}\delta_{gs}\right)
   [\left(\delta_{hp}\delta_{fr}-\delta_{hr}\delta_{fp}\right)
        \left(\epsilon^e_h+\epsilon^e_f\right)
   \right.\\\nonumber &+&\left.
        \frac{1}{2}\sum_{mtlk}\left(\delta_{ft}\delta_{hm}-\delta_{fm}\delta_{ht}\right)
        \left(\delta_{kp}\delta_{lr}-\delta_{lp}\delta_{kr}\right)
        V^{eeee}_{mtlk}]
    \right.\\\nonumber &-& \left.
    \left(\delta_{hp}\delta_{fr}-\delta_{fp}\delta_{hr}\right)
   [\left(\delta_{gs}\delta_{eq}-\delta_{es}\delta_{gq}\right)
        \left(\epsilon^h_s+\epsilon^h_q\right)
        \right.\\\nonumber &-&\left.
        \frac{1}{2}\sum_{mtlk}\left(\delta_{tq}\delta_{ms}-\delta_{ts}\delta_{mq}\right)
        \left(\delta_{gl}\delta_{eq}-\delta_{gk}\delta_{el}\right)
        V^{hhhh}_{mtlk}]
    \right.\\\nonumber&+&\left.
    \sum_{mtlk}[\left(\delta_{tq}\left(\delta_{gl}\delta_{es}-\delta_{gs}\delta_{el}\right)
    -\delta_{ts}\left(\delta_{gl}\delta_{eq}-\delta_{gq}\delta_{el}\right) \right)
    \right.\\\nonumber&\times&\left.
    \left(\delta_{kp}\left(\delta_{hm}\delta_{fr}-\delta_{hr}\delta_{fm}\right)
    -\delta_{kr}\left(\delta_{hm}\delta_{fp}-\delta_{hp}\delta_{fm}\right)\right)
      \right. \nonumber \\
     &\times& (V^{eheh}_{mtkl}-V^{ehhe}_{mtlk})] \nonumber \\\nonumber
    &-&\left.
    \left(\delta_{gq}\delta_{es}-\delta_{eq}\delta_{gs}\right)
   \left(\delta_{hp}\delta_{fr}-\delta_{hr}\delta_{fp}\right)\hbar\omega^{xx}_{a}\right\}U^{xx}_{a,pqrs}=0,\;\;\;\; 
\end{eqnarray}
respectively. Using the transformation matrix defined by the secular equations Eqs.~(\ref{U1}) and (\ref{U2}) the Coulomb matrix element entering Eq.~(\ref{HW}) becomes
\begin{eqnarray}	\label{WW12}
   	V^{x,xx}_{a,n}&=&
	\\\nonumber&~&
	\sum_{pq}\sum_{rstv}\left(U^x_{a;pq}\right)^*
   	\left[(V^{ehee}_{pstr}-V^{ehee}_{psrt})\delta_{qv}\right.
	\nonumber \\
	&-&\left. (V^{ehee}_{pvtr}-V^{ehee}_{pvrt})\delta_{qs}\right]U^{xx}_{rstv;n}
   \nonumber \\ &+&
   	\sum_{pq}\sum_{rstv}\left(U^x_{a;pq}\right)^*
   	\left[(V^{hhhe}_{vsqr}-V^{hhhe}_{svqr})\delta_{pt}\right. \nonumber \\
	&-&\left. (V^{hhhe}_{svqt}-V^{hhhe}_{vsqt})\delta_{pr}\right]U^{xx}_{rstv;n}
   	\nonumber \\ \label{V02}
	V^{0,xx}_{0,n}&=&
	\sum_{rstv}\left[V^{hhee}_{vsrt}-V^{hhee}_{vstr}\right]U^{xx}_{rstv;n}.
\end{eqnarray}
In Eqs.~(\ref{U1-opr})-(\ref{WW12}) indices $a$ and $b$ denote the single- and bi-exciton eigenstates, respectively. The rest of the indices describe carriers states.

\section{The exciton-phonon interaction Hamiltonian}
\label{RxxH}

The many-body Hamiltonian accounting for the linear electron-phonon interaction in the electron-hole representation is\cite{axt98}
\begin{eqnarray}\label{HMB_ep-int}
    \hat{\cal H}_{int} &=& \sum_{mn\alpha}f^{ee}_{mn;\alpha}q_\alpha c_m^\dag c_n
    -\sum_{mn\alpha}f^{hh}_{mn;\alpha}q_\alpha d_m^\dag d_n
    \\\nonumber
     &+&\sum_{nm\alpha}f^{he}_{mn;\alpha}q_\alpha d_m c_n
     + \sum_{mn\alpha} f^{eh}_{mn;\alpha} q_\alpha c_m^\dag d_n^\dag,
\end{eqnarray}
where the phonon normal modes are ${\bf q}=\{q_1, \dots q_\alpha\dots q_{N_{ph}}\}$, and the coupling constants are the following matrix elements
\begin{eqnarray}
 \label{kappee}
    f_{mn;\alpha}^{rs} &=& \langle\psi_m^r|\hat F_\alpha|\psi_n^s\rangle, \;\;\;r,s=e,h;
\end{eqnarray}
of the force operator $\hat F_\alpha$ averaged over the Hartree-Fock electron and hole wave functions.  In this Hamiltonian, we dropped the term, $\sum_{n\alpha}f^{hh}_{nn;\alpha}q_{\alpha}$, which has no contribution to the processes under consideration.

Modifying the multi-exciton basis set given by Eq.~(\ref{multi-X-bas}), the electron-phonon interaction Hamiltonian~(\ref{HMB_ep-int}) can be projected on single- and bi-exciton states resulting in Eq.~(\ref{Hxp-int}) where the intraband exciton-phonon coupling matrix elements are
\begin{eqnarray}\label{Y-X}
 Y^{x}_{ab;\alpha}&=& \sum_{pq=e}\sum_{r=h}U^x_{a;pr}f^{ee}_{pq;\alpha}U^x_{b;qr}
 \\\nonumber&-&
 \sum_{pq=h}\sum_{r=e} U^x_{a;rp}f^{hh}_{pq;\alpha} U^x_{b;rq},
 \\\label{Y-XX}
 Y^{xx}_{mn;\alpha}&=& \sum_{kgpr=e}\sum_{hfqs=h}U^{xx}_{m;khgf} U^{xx}_{n;pqrs}
 \\\nonumber&\times&
 \left(\delta_{hq}\delta_{fs}-\delta_{fq}\delta_{hs}\right)
 \\\nonumber&\times&
 \left(\delta_{gr}f^{ee}_{kp;\alpha} - \delta_{gp}f^{ee}_{kr;\alpha}
 -\delta_{kr}f^{ee}_{ga;\alpha}+ \delta_{ka}f^{ee}_{gr;\alpha}\right)
 \\\nonumber
 &-&\sum_{kgpr=e}\sum_{hfqs=h}U^{xx}_{m;khgf} U^{xx}_{n;pqrs}
 \\\nonumber&\times&
 \left(\delta_{kp}\delta_{gr}-\delta_{gp}\delta_{kr}\right)
 \\\nonumber&\times&
 \left(\delta_{fs}f^{hh}_{hq;\alpha} -\delta_{hs}f^{hh}_{fq;\alpha}
 -\delta_{fq}f^{hh}_{hs;\alpha}+\delta_{hq}f^{hh}_{fs;\alpha}\right).
\end{eqnarray}
The interband exciton-phonon matrix element also entering Eq.~(\ref{Hxp-int}) is
\begin{eqnarray}
 \label{Y-X-XX}
 Y^{x,xx}_{am;\alpha}&=&\sum_{gpr=e}\sum_{fqs=h}
 U^x_{a;gf}U^{xx}_{m;pqrs}
 f^{he}_{mn;\alpha}
 \\\nonumber&\times&
 \left(\delta_{fs}\delta_{gr}f^{he}_{qp;\alpha}-\delta_{fs}\delta_{gp}f^{he}_{qr;\alpha}
 \right.\\\nonumber &-&\left.
 \delta_{fq}\delta_{gr}f^{he}_{sp;\alpha}+\delta_{fg}\delta_{gp}f^{he}_{sr;\alpha}\right).
\end{eqnarray}
Here, the transformation matrices $U^x$ and $U^{xx}$ can be calculated according to Eqs.~(\ref{U1}) and (\ref{U2}), respectively.

\section{The exciton-optical field interaction Hamiltonian}
\label{eofh}

The many-body Hamiltonian in the Hartree-Fock orbital representation describing the interaction with the time-dependent optical field $E(t)$ is:\cite{axt98}
\begin{eqnarray}\label{H-opt-mb}
 \hat{\cal H}_{opt}(t) &=& -E(t)\sum_{mn}P^{ee}_{mn} c_m^\dag c_n
 \\\nonumber&+&
 	E(t)\sum_{mn}P^{hh}_{mn} d_m^\dag d_n
    \\\nonumber&-&
    E(t)\sum_{nm}P^{he}_{mn} d_m c_n
    \\\nonumber&-&
     E(t)\sum_{mn} P^{eh}_{mn} c_m^\dag d_n^\dag,
\end{eqnarray}
where the transition dipole moments are matrix elements
\begin{eqnarray}
 \label{kappee}
    P_{mn}^{rs} &=& \langle\psi_m^r|{\bf \hat e \cdot \hat d}|\psi_n^s\rangle, \;\;\;r,s=e,h;
\end{eqnarray}
of the dipole moment operator ${\bf\hat d}$ projected onto the field polarization direction, ${\bf\hat e}$, and further averaged over the Hartree-Fock electron and hole wave functions. This Hamiltonian has exactly the same structure as the electron-phonon coupling Hamiltonian (Eq.~(\ref{HMB_ep-int})) discussed in Appendix~\ref{RxxH}. Therefore, in direct analogy with Eq.~(\ref{Hxp-int}), one can immediately recast the former Hamiltonian to the single and bi-exciton state representation given by Eq.~(\ref{H-opt}). The intraband single-exciton (bi-exciton) transition dipoles $\mu^{x}_{ab}$ ($\mu^{xx}_{mn}$) entering Eq.~(\ref{H-opt}) can be determined by replacing $Y^{x}_{ab;\alpha}$ ($Y^{xx}_{nm;\alpha}$) in the l.h.s. of Eq.~(\ref{Y-X}) (Eq.~(\ref{Y-XX})) by $\mu^{x}_{ab}$ ($\mu^{xx}_{nm}$). Determination of the interband transition dipoles requires the replacement of $Y^{x,xx}_{am;\alpha}$ in the l.h.s. of Eq.~(\ref{Y-X-XX}) by $\mu^{x,xx}_{am}$.  Also all $f^{rs}_{pq;\alpha}$ ($r,s=e,h$) in the r.h.s.\@ of Eqs.~(\ref{Y-X})--(\ref{Y-X-XX}) should be replaced by $P^{rs}_{ab}$ ($r,s=e,h$).

\section{Coherent Superposition Model}
\label{cspm}

Coherent Superposition Model is the limit of the Exciton Scattering Model including only two states $|x_{c}\rangle$  and $|xx_{c}\rangle$ coupled by Coulomb matrix element $V^{x,xx}$ and two uncoupled states states $|x_{u}\rangle$ and $|xx_{u}\rangle$ characterized by the following Hamiltonian: 
\begin{eqnarray}
    \label{H-cw}
    \hat H &+& \sum_{a=c,u}\left(|x_{a}\rangle \hbar\omega^{x}_a \langle x_{a}|
    		+|xx_{a}\rangle \hbar\omega^{xx}_a \langle xx_{a}|\right)
\\\nonumber&+&
              |x_{c}\rangle V^{x,xx} \langle xx_{c}|+|xx_{c}\rangle V^{xx,x}\langle x_{c}|.
\end{eqnarray}

According to Eqs.~(\ref{TF-intra}) and (\ref{TF-inter}), the scattering matrix elements for the coupled states are
\begin{eqnarray}
    \label{TX}
    T^x_c(\omega)&=&\left(\frac{V^{x,xx}}{\hbar}\right)^2
        \frac{\left(\omega-\omega_c^{x}\right)}
        {i\left(\omega-\omega_{+}\right)\left(\omega-\omega_{-}\right)},
   \\\label{TXX}
    T^{xx}_c(\omega)&=&\left(\frac{V^{x,xx}}{\hbar}\right)^2
        \frac{\left(\omega-\omega_c^{xx}\right)}
        {i\left(\omega-\omega_{+}\right)\left(\omega-\omega_{-}\right)},
   \\\label{TX-XX}
    T^{x,xx}_c(\omega)&=&\frac{V^{x,xx}}{\hbar}
        \frac{\left(\omega-\omega_c^{x}\right)
        \left(\omega-\omega_c^{xx}\right)}
        {i\left(\omega-\omega_{+}\right)\left(\omega-\omega_{-}\right)},
\end{eqnarray}
where we denote the eigenenergies (quasiparticle energies) as
\begin{eqnarray}
    \label{wpm}
    \omega_{\pm}=\frac{\omega_c^{x}+\omega_c^{xx}}{2}\pm
    \sqrt{\left(\frac{\omega^{x}_c-\omega^{xx}_c}{2}\right)^2
    +\left(\frac{V^{x,xx}}{\hbar}\right)^2}.
\end{eqnarray}

Following the procedure in Sec.~\ref{esmd}, one finds that the corresponding time-dependent Green function components for coupled states are 
\begin{eqnarray}
\label{GtX-sc}
    G^{x}_c(t)&=& \Lambda^{x}_+ e^{-i\tilde\omega_{+}t}
            +\Lambda^{x}_- e^{-i\tilde\omega_{-}t}
\\\label{GtXX-sc}
    G^{xx}_c(t)&=& \Lambda^{xx}_+ e^{-i\tilde\omega_{+}t}
            +\Lambda^{xx}_- e^{-i\tilde\omega_{-}t}
    \\\label{GtInt-sc}
    G^{x,xx}_c(t)&=& \Lambda^{x,xx}_+ e^{-i\tilde\omega_{+}t}
            +\Lambda^{x,xx}_- e^{-i\tilde\omega_{-}t},
\end{eqnarray}
which depend on the transition amplitudes 
\begin{eqnarray}
\label{LmX}
    \Lambda^x_{\pm}&=& \pm\frac{\left(\omega_\pm-\omega_c^{xx}\right)}
                {\left(\omega_+-\omega_-\right)},
\\\label{LmXX}
    \Lambda^{xx}_{\pm}&=& \pm\frac{\left(\omega_\pm-\omega_c^{x}\right)}
                {\left(\omega_+-\omega_-\right)},    
\\\label{LmInt}
    \Lambda^{x,xx}_\pm&=& \pm\frac{V^{x,xx}}
            {\hbar\left(\omega_{+}-\omega_{-}\right)}.
\end{eqnarray}
Note that these quantities are real, since the Coulomb coupling significantly exceeds the level broadening. 

To find the single- and bi-exciton populations associated with the excitation of the quasiparticle populations, we use the following relaxation equations for populations in the quasiparticle representation (Fig.~\ref{Fig-csm}~(b)):
\begin{eqnarray}\label{Lpop}
	\dot\rho_\pm &=& -\left(\Gamma_{x,\pm}+\Gamma_{xx,\pm}\right)\rho_\pm
\\\nonumber
	\dot\rho^x_u &=& \Gamma_{x,+}\rho_+ + \Gamma_{x,-}\rho_-
\\\nonumber
	\dot\rho^{xx}_u &=& \Gamma_{xx,+}\rho_+ + \Gamma_{xx,-}\rho_-.
\end{eqnarray}
The non-vanishing (for $\tau>0$) components of the Liouville space Green function (Eqs.~(\ref{Lpop})) are:
\begin{eqnarray}\label{LG}
	\bar{\cal G}_{\pm;\pm}(\tau) &=& e^{-\left(\Gamma_{x,\pm}+\Gamma_{xx,\pm}\right)\tau},
\\\nonumber
	\bar{\cal G}^x_{u;\pm}(\tau) &=&\frac{\Gamma_{x,\pm}}{\Gamma_{x,\pm}+\Gamma_{xx,\pm}}
												\left(1-\bar{\cal G}_{\pm;\pm}(\tau)\right),
\\\nonumber
	\bar{\cal G}^{xx}_{u;\pm}(\tau) &=& \frac{\Gamma_{xx,\pm}}{\Gamma_{x,\pm}+\Gamma_{xx,\pm}}
												\left(1-\bar{\cal G}_{\pm;\pm}(\tau)\right).
\end{eqnarray}
where, the population relaxation rates $\Gamma_{x,\pm}$ and $\Gamma_{xx,\pm}$ are given by Eqs.~(\ref{prx-cw}) and (\ref{prxx-cw}), respectively. For the sake of simplicity, we dropped the population transfer rates between the quasiparticle states. 

The time-dependent single and bi-exciton populations due to the excitation of quasiparticle populations are calculated according Eqs.~(\ref{mu-eigs}) and (\ref{rhop-qprt}) where  Eqs.~(\ref{LmX})--(\ref{LmInt}) and (\ref{LG}) are substituted. For the coupled states these populations are
\begin{eqnarray}\label{nxc}
   n^x_c (\tau) &=& \mu^2\sum_{\bar\xi=\pm}\bar{\cal G}_{\bar\xi,\bar\xi}(\tau)
   [{\Lambda^x_{\bar\xi}}]^2{\cal I}(\omega_{\bar\xi}-\omega_{pm}),
 \\\label{nxxc}
   n^{xx}_c (\tau) &=& \mu^2\sum_{\bar\xi=\pm}\bar{\cal G}_{\bar\xi,\bar\xi}(\tau)
   [{\Lambda^{x,xx}_{\bar\xi}}]^2{\cal I}(\omega_{\bar\xi}-\omega_{pm}),\;\;\;
\end{eqnarray}
and for the uncoupled states
\begin{eqnarray}\label{nxu}	
	n^x_u (\tau) &=&\mu^2\sum_{\bar\xi=\pm}\bar{\cal G}^x_{u,\bar\xi}(\tau)
						{\Lambda^x_{\bar\xi}}{\cal I}(\omega_{\bar\xi}-\omega_{pm}),
 \\\label{nxxu}
   n^{xx}_u (\tau) &=&\mu^2\sum_{\bar\xi=\pm}\bar{\cal G}^{xx}_{u,\bar\xi}(\tau)
   						{\Lambda^{x}_{\bar\xi}}{\cal I}(\omega_{\bar\xi}-\omega_{pm}).
\end{eqnarray}
Here, the pulse self-convolution function is defined by Eq.~(\ref{Ip-eigs}), and the population Green functions by Eqs.~(\ref{LG}).

The Liouville equation for the coupled state coherence in the quasiparticle basis set is:
\begin{eqnarray}\label{Lcoh}
	\dot\rho_{+-} &=& -i\tilde\omega_{+-}\rho_{+-},
\end{eqnarray}
where $\tilde\omega_{+-}=\omega_{+}-\omega_{-}-i\gamma_{+-}$ contains $\omega_\pm$ determined by Eq.~(\ref{wpm}) and the pure dephasing  rate $\gamma_{+-}$ which can be explicitly found by using Eq.~(\ref{dphas-rate}) and the following quasiparticle-phonon couplings: 
\begin{eqnarray}\label{dphas-cw}
    Y_{\pm,\alpha} &=& Y^x_{c;\alpha}\Lambda^x_\pm+ 2Y^{x,xx}_{cc;\alpha}\Lambda^{x,xx}_\pm+Y^{xx}_{c;\alpha}\Lambda^{xx}_\pm,
\end{eqnarray}
where $Y^x_{c;\alpha}$, $Y^{x,xx}_{uc;\alpha}$, $Y^{xx,x}_{cc;\alpha}$, $Y^{xx}_{c;\alpha}$ are the components of the exciton states coupled to the phonon mode, $\alpha$.

According to Eq.~(\ref{rhoc-qprt}), the contribution of the quasiparticle coherences to the coupled single- and bi-exciton populations are:
\begin{eqnarray}\label{cxc}
   c^x_c (\tau) &=& 2\mu^2 {\Lambda^x_{+}}{\Lambda^x_{-}}
   Re~\left\{e^{-i\tilde\omega_{+-}\tau}
 \right.\\\nonumber&\times&\left.
   {\cal I}(\omega_+-\omega_{pm};\omega_--\omega_{pm})\right\}
\\\label{cxxc}
   c^{xx}_c (\tau) &=& 2\mu^2 
   {\Lambda^{xx,x}_{+}}{\Lambda^{xx,x}_{-}}
   Re~\left\{e^{-i\tilde\omega_{+-}\tau}
 \right.\\\nonumber&\times&\left.
   {\cal I}(\omega_+-\omega_{pm};\omega_--\omega_{pm})\right\},
\end{eqnarray}
where the pulse self-convolution function is given by Eq.~(\ref{Ic-eigs}). By taking into account that ${\Lambda^{xx,x}_{+}}{\Lambda^{xx,x}_{-}}=-{\Lambda^x_{+}}{\Lambda^x_{-}}$, one finds that $c^x_c (\tau)=-c^{xx}_c (\tau)$. Obviously, $c^x_u (\tau)=c^{xx}_u (\tau) =0$.

\bibliographystyle{prsty}

\begin{thebibliography}{10}

\bibitem{kolodinski93}
S. Kolodinski, J. Werner, T. Wittchen, and H. Queisser, Appl.~Phys.~Lett. {\bf
  63},  2405  (1993).

\bibitem{nozik02}
A.~J. Nozik, Physica~E {\bf 14},  115   (2002).

\bibitem{hanna06}
M. Hanna and A. Nozik, J.~Appl.~Phys. {\bf 100},  074510  (2006).

\bibitem{klimov06}
V. Klimov, Appl.~Phys.~Lett. {\bf 89},  123118  (2006).

\bibitem{beard08a}
M.~C. Beard and R.~J. Ellingson, Laser~\&~Photon.~Rev. {\bf 2},  377  (2008).

\bibitem{nozik08}
A.~J. Nozik, Chem.~Phys.~Lett. {\bf 457},  3  (2008).

\bibitem{luther08}
J.~M. Luther, M. Law, M. C. Beard, Q. Song, M. O. Reese, R.~J. Ellingson, and A.~J. Nozik, Nano Lett. {\bf 8},  3488  (2008).

\bibitem{koc57}
S. Koc, Czech.~J.~Phys. {\bf 7},  91  (1957).

\bibitem{smith58}
A. Smith and D. Dutton, J. Opt. Soc. Am. {\bf 48},  1007   (1958).

\bibitem{vavilov59}
V. Vavilov, J.~Phys.~Chem.~Solids {\bf 8},  223  (1959).

\bibitem{tauc59}
J. Tauc, J.~Phys.~Chem.~Solids {\bf 8},  219  (1959).

\bibitem{cristensen76}
O. Christensen, J.~Appl.~Phys. {\bf 47},  689  (1976).

\bibitem{pijpers09}
J.~J.~H. Pijpers, R. Ulbricht, K. J. Tielrooij, A. Osherov, Y. Golan, C. Delerue, G. Allan, and M. Bonn, Nature~Phys.  {\bf 5},  811  (2009).

\bibitem{shockley61a}
W. Shockley, Czech.~J.~Phys. {\bf 11},  81  (1961).

\bibitem{kane67}
E. Kane, Phys.~Rev. {\bf 159},  624  (1967).

\bibitem{antoncik78}
E. Antoncik and N. Gaur, J.~Phys.~D: Solid State Phys. {\bf 11},  735  (1978).

\bibitem{wolf98}
M. Wolf, R. Brendel, J.~H. Werner, and H.~J. Queisser, J.~Appl.~Phys. {\bf 83},
   4213  (1998).

\bibitem{landsberg91}
P.~T. Landsberg, {\em Recombination in Semiconductors} (Cambridge University
  Press, Cambridge, 1991).

\bibitem{harrison99}
D. Harrison, R.~A. Abram, and S. Brand, J.~Appl.~Phys. {\bf 85},  8186  (1999).

\bibitem{klein68}
C.~A. Klein, J.~Appl.~Phys. {\bf 39},  2029   (1968).

\bibitem{alig75}
R.~C. Alig and S. Bloom, Phys.~Rev.~Lett. {\bf 35},  1522   (1975).

\bibitem{chepic90}
D.~I. Chepic, A.~L. Efros, A.~I. Ekimov, M.~G. Vanov, V.~A. Kharchenko, 
I.~A. Kudriavtsev, and T.~V. Yazeva, J.~Lumin. {\bf 47},  113   (1990).

\bibitem{nozik01}
A.~J. Nozik, Annu.~Rev.~Phys.~Chem. {\bf 52},  193   (2001).

\bibitem{klimov00}
V.~I. Klimov, A.~A. Mikhailovsky, S. Xu, A. Malko, J.~A. Hollingsworth, 
	C.~A. Leatherdale, H.~J. Eisler, and M.~G.  Bawendi, 
	Science {\bf 290},  314   (2000).

\bibitem{schaller04}
R.~D. Schaller and V.~I. Klimov, Phys.~Rev.~Lett. {\bf 92},  186601   (2004).

\bibitem{ellingson05}
R.~J. Ellingson, M.~C. Beard, J.~C. Johnson, P.~R. Yu, O.~I. Micic, A.~J. Nozik,
 A. Shabaev, and A.~L. Efros, Nano Lett. {\bf 5},  865   (2005).

\bibitem{schaller05}
R.~D. Schaller, V.~M. Agranovich, and V.~I. Klimov, Nature~Phys. {\bf 1},  189
   (2005).

\bibitem{schaller05a}
R.~D. Schaller, M.~A. Petruska, and V.~I. Klimov, Appl.~Phys.~Lett. {\bf 87},
  253102   (2005).

\bibitem{schaller06}
R.~D. Schaller, M. Sykora, J.~M. Pietryga, and V.~I. Klimov, Nano Lett. {\bf
  6},  424   (2006).

\bibitem{schaller06a}
R.~D. Schaller and V.~I. Klimov, Phys.~Rev.~Lett. {\bf 96},  097402   (2006).

\bibitem{schaller06b}
R.~D. Schaller, M. Sykora, S. Jeong, and V.~I. Klimov, J.~Phys.~Chem.~B {\bf
  110},  25332   (2006).

\bibitem{murphy06}
J.~E. Murphy, M.~C. Beard, A.~G. Norman, S.~P. Ahrenkiel, J.~C. Johnson, P.~R. Yu,
 O.~I. Micic, R.~J. Ellingson, and A.~J. Nozik, J.~Am.~Chem.~Soc. {\bf 128},  3241   (2006).

\bibitem{schaller07}
R.~D. Schaller, J.~M. Pietryga, and V.~I. Klimov, Nano Lett. {\bf 7},  3469
  (2007).

\bibitem{beard07}
M.~C. Beard, K.~P. Knutsen, P.~R. Yu, J.~M. Luther, Q. Song, W.~K. Metzger, R.~J. Ellingson,
 and A.~J. Nozik, Nano Lett. {\bf 7},  2506   (2007).

\bibitem{pijpers07}
J.~J.~H. Pijpers, E. Hendry, M.~T.~W. Milder, R. Fanciulli, J. Savolainen, J.~L. Herek, 
D. Vanmaekelbergh, S. Ruhman, D. Mocatta, D. Oron, A. Aharoni, U. Banin, and M. Bonn, 
J.~Phys.~Chem.~C {\bf 111},  4146   (2007).

\bibitem{nair07}
G. Nair and M.~G. Bawendi, Phys.~Rev.~B {\bf 76},  081304(R)  (2007).

\bibitem{benlulu08}
M. Ben-Lulu, D. Mocatta,  M. Bonn, U. Banin, S. Ruhman, Nano Lett. {\bf 8},  1207  (2008).

\bibitem{pijpers08}
J.~J.~H. Pijpers, E. Hendry, M.~T.~W. Milder, R. Fanciulli, J. Savolainen, and J.~L. Herek, D. Vanmaekelbergh
 S. Ruhman, D. Mocatta, D. Oron, A. Aharoni, U. Banin, and M. Bonn, J.~Phys.~Chem.~C {\bf 112},  4783  (2008).

\bibitem{mcguire08}
J.~A. Mcguire, J. Joo, J.~M. Pietryga, R.~D. Schaller, and V.~I. Klimov, Acc.~Chem.~Res. {\bf 41},  1810  (2008).

\bibitem{trinh08}
M.~T. Trinh, A.~J. Houtepen, J.~M. Schins, T. Hanrath, J. Piris, W. Knulst, A.~P.~L.~M. Goossens, 
and L.~D.~A. Siebbeles, Nano Lett. {\bf 8},  1713  (2008).

\bibitem{nair08}
G. Nair, S.~M. Geyer, L.-Y. Chang, and M.~G. Bawendi, Phys.~Rev.~B {\bf 78},
  125325  (2008).

\bibitem{ji09}
M. Ji, S. Park, S.~T. Connor, T. Mokari, Y. Cui, and K.~J. Gaffney, Nano Lett. {\bf 9},  1217  (2009).

\bibitem{franceschetti08}
A. Franceschetti and Y. Zhang, Phys.~Rev.~Lett. {\bf 100},  136805  (2008).

\bibitem{beard09}
M.~C. Beard, A.~G. Midgett, M. Law, O.~E. Semonin, R.~J. Ellingson, and A.~J. Nozik, 
Nano Lett. {\bf 9},  836  (2009).

\bibitem{kilina09a}
S. Kilina, S. Ivanov, and S. Tretiak, J.~Am.~Chem.~Soc. {\bf 131},  7717
  (2009).

\bibitem{shabaev06}
A. Shabaev, A.~L. Efros, and A.~J. Nozik, Nano Lett. {\bf 6},  2856   (2006).

\bibitem{rupasov07}
V.~I. Rupasov and V.~I. Klimov, Phys.~Rev.~B {\bf 76},  125321  (2007).

\bibitem{isborn08}
C.~M. Isborn, S.~V. Kilina, X. Li, and O.~V. Prezhdo, J.~Phys.~Chem.~C {\bf
  112},  18291–18294  (2008).

\bibitem{franceschetti06}
A. Franceschetti, J.~M. An, and A. Zunger, Nano Lett. {\bf 6},  2191   (2006).

\bibitem{allan06}
G. Allan and C. Delerue, Phys.~Rev.~B {\bf 73},  205423   (2006).

\bibitem{allan08}
G. Allan and C. Delerue, Phys.~Rev.~B {\bf 77},  125340  (2008).

\bibitem{luo08}
J.-W. Luo, A. Franceschetti, and A. Zunger, Nano Lett. {\bf 8},  3174  (2008).

\bibitem{allan09}
G. Allan and C. Delerue, Phys.~Rev.~B {\bf 79},  195324  (2009).

\bibitem{rabani08}
E. Rabani and R. Baer, Nano Lett. {\bf 8},  4488  (2008).

\bibitem{axt98}
V.~M. Axt and S. Mukamel, Rev.~Mod.~Phys. {\bf 70},  145   (1998).

\bibitem{mukamel_book}
S. Mukamel, {\em Principles of Nonlinear Optical Spectroscopy} (Oxford
  University Press, Oxford, 1995).

\bibitem{economou_greenf}
E.~N. Economou, {\em Green's functions in Quantum Physics} (Springer-Verlag,
  New York, 1983).

\bibitem{guyotsionnest05}
P. Guyot-Sionnest, B. Wehrenberg, and D. Yu, J.~Chem.~Phys. {\bf 123},  074709
  (2005).

\bibitem{pandey08}
A. Pandey and P. Guyot-Sionnest, Science {\bf 322},  929  (2008).

\bibitem{kilina07}
S.~V. Kilina, C.~F. Craig, D.~S. Kilin, and O.~V. Prezhdo, J.~Phys.~Chem.~C
  {\bf 111},  4871  (2007).

\bibitem{kilina09}
S.~V. Kilina, D.~S. Kilin, and O.~V. Prezhdo, ACS Nano {\bf 3},  93  (2009).

\bibitem{kubo95}
R. Kubo, M. Toda, and N. Hashitsume, {\em Statistical Physics II}, No.~31 in
  {\em Solid-state sciences}, 2nd ed. (Springer, Berlin, 1995).

\bibitem{vankampen92}
N.~G. van Kampen, {\em Stochastic Processes in Physics and Chemistry}
  (North-Holland, Amsterdam, 1992).

\bibitem{dahlbom00}
M. Dahlbom, T. Minami, V. Chernyak, T. Pullerits, V. Sundstrom, and S. Mukamel, 
J.~Phys.~Chem.~B {\bf 104},  3976   (2000).

\bibitem{ernst90}
R.~R. Ernst, G. Bodenhausen, and A. Wokaun, {\em Principles of Nuclear Magnetic
  Resonance in One and Two Dimensions} (Clarendon Press, Oxford, 1990).

\bibitem{Zwanzig_NSM}
R. Zwanzig, {\em Nonequilibrium Statistical Mechanics} (Oxford University
  Press, New York, 2001).

\bibitem{kamisaka06}
H. Kamisaka, S.~V. Kilina, K. Yamashita, and O.~V. Prezhdo, Nano Lett. {\bf 6},
   2295   (2006).

\bibitem{kamisaka08}
H. Kamisaka, S.~V. Kilina, K. Yamashita, and O.~V. Prezhdo, J.~Phys.~Chem.~C {\bf
  112},  7800  (2008).
\bibitem{ftn-01}
The pump envelope and spatial phases do not contribute to the population dynamics and therefore are dropped.

\bibitem{ftn-02}
Using this representation for the transition amplitude, Eqs.~(\ref{Yqp-d}) and (\ref{Yqp-o}) can be derived by straight forward transformation between the exciton and quasiparticle basis sets using the following relationship $\langle l|\bar\xi\rangle\langle\bar\xi|r\rangle = \bar\Lambda_{lr}(\omega_{\bar\xi})$ discussed in Ref.~[\onlinecite{economou_greenf}]

\bibitem{ftn-03}
Eqs.~(\ref{rho-eq})--(\ref{LmbdX-eq}) are obtained in the second order perturbation theory where the small parameter is $V^{x,xx}_{0,k}/E_g$. We also used the fact that in semiconductors NCs $E_g\gg k_B T$.

\bibitem{ftn-04}
The poles are assumed to be of the first order which is the general situation for arbitrary Coulomb coupling. Two poles can coincide if we use the second order expansion of Eq.~(\ref{GT-rep}). This results in the renormalization of the energy in the exponential but the form of Eq.~(\ref{G-xres}) will still be the same.

\bibitem{ftn-05}
Deriving Eqs.~(\ref{rhoc-qprt}) and (\ref{mu-eigs}), we used the following relationship between the components of the equilibrium density matrix and transition amplitude: $\Lambda_{lr}(\omega_0) = \bar\rho_{lr}$.

\bibitem{ftn-06}
Note that in Eq.~(\ref{rhop-qprt}) for $n_s(0)$, the population Green function becomes $\bar{\cal G}_{\bar\zeta,\bar\xi}(0)=\delta_{\bar\zeta,\bar\xi}$, and associated prefactor simplifies to $\sum_{\bar\zeta}\left[\bar\Lambda_{ss}(\omega_{\bar\zeta})\bar{\cal G}_{\bar\zeta,\bar\xi}(0)\bar\Lambda_{lr}(\omega_{\bar\xi})\right] = \delta_{sl}\delta_{sr}$.

\bibitem{ftn-07}
Formally, the expression for the interband population transfer rate $\Gamma^{x,xx}_{ak} = \hbar^{-2}\sum_{\alpha\alpha^{'}} Y^{x,xx}_{ak;\alpha}Y^{xx,x}_{ka;\alpha^{'}} C_{\alpha\alpha^{'}}(\omega^{x}_{a}-\omega^{xx}_{k})$ can be introduced. However, in this expression $Y^{xx,x}_{ka;\alpha^{'}}\neq 0$ only for the states which have $\hbar(\omega^{xx}_{k}-\omega^{x}_{a})>E_g$. Since $E_g$ significantly exceeds the phonon bath spectral widths, $\Gamma^{x,xx}_{ak} = 0$.

\bibitem{ftn-08}
This follows from the observation that for $V^{x,xx}\rightarrow 0$, $\Lambda^x_{+}=\Lambda^{xx}_{-} = 1/2$ and  $\Lambda^x_{-}=\Lambda^{xx}_{+} = 0$.

\bibitem{ftn-09}
It is easy to show that for small $\tau_{pm}$ the Gaussian pulse envelope can be approximated as ${\cal E}_{pm}(t) = {\cal E}^{(0)}_{pm} e^{-t^2/2\tau^2_{pm}}\approx\sqrt{2\pi}\tau{\cal E}^{(0)}_{pm}\delta(t)$ and further substituted into Eqs.~(\ref{Ic-eigs})~and~(\ref{Ip-eigs}) to get the pulse self-convolution functions.

\bibitem{ftn-10}
The equalities require that $C_\alpha(\omega)$ is a constant for $\omega$ varying within the range defined by $V^{x,xx}$.

\bibitem{ftn-11}
This expression can be rigorously derived by using the Gaussian form for the envelope function ${\cal E}_{pm}(t) = {\cal E}^{(0)}_{pm} e^{-t^2/2\tau^2_{pm}}$. Another way to derive Eq.~(\ref{ICW}) for a generic pulse envelope is to assume that ${\cal E}^{(0)}_{pm}=$const. during the coherence integral over $t_1$, and that the pulse auto-correlation function $\int_{-\infty}^{\infty}dt^{'}{\cal E}_{pm}(t^{'}){\cal E}_{pm}(t^{'}-t_1)\sim \tau_{pm}~{{\cal E}^{(0)}_{pm}}^2$.

\bibitem{ftn-12}
The self-energy contributions to the energy are dropped because they correspond to fourth-order corrections.

\bibitem{ftn-13}
Notice that this equation does not account for the exciton filling factor effects since no more than a single- or bi-exicton states are excited per NC. In fact, the filling factors for the carriers forming the excitons are already accounted for by the exciton definition and enter into the Redfield equation implicitly.

\end{thebibliography}

\end{document}